\documentclass[apj]{emulateapj}
\usepackage{graphicx}
\usepackage{amsmath,amssymb}
\usepackage{mathptmx}
\usepackage[english]{babel}
\usepackage{booktabs}
\usepackage{multirow}
\usepackage{natbib}
\usepackage[breaklinks,colorlinks,
   urlcolor=blue,citecolor=blue,linkcolor=blue]{hyperref}
\usepackage{color}
\usepackage{verbatim}

\newcommand{\beq}	{\begin{equation}}
\newcommand{\eeq}	{\end{equation}}
\newcommand{\beqs}	{\begin{displaymath}}
\newcommand{\eeqs}	{\end{displaymath}}
\newcommand{\beqa}{\begin{eqnarray}}
\newcommand{\eeqa}{\end{eqnarray}}
\newcommand{\beqas}	{\begin{eqnarray*}}
\newcommand{\eeqas}	{\end{eqnarray*}}

\definecolor{red}{rgb}{0.9,0,0}
\definecolor{blue}{rgb}{0,0,0.9}
\definecolor{dgreen}{rgb}{0,0.6,0}

\newcommand{\ovi}{O{\sc~VI}}
\newcommand{\ovii}{O{\sc~VII}}
\newcommand{\oviii}{O{\sc~VIII}}
\newcommand{\nv}{N{\sc~V}}
\newcommand{\neviii}{Ne{\sc~VIII}}
\newcommand{\mgx}{Mg{\sc~X}}
\def\novi{\ifmmode {{\rm N_{\rm OVI}}} \else ${\rm N_{\rm OVI}}$\fi}
\def\novii{\ifmmode {{\rm N_{\rm OVII}}} \else ${\rm N_{\rm OVII}}$\fi}
\def\noviii{\ifmmode {{\rm N_{\rm OVIII}}} \else ${\rm N_{\rm OVIII}}$\fi}
\def\nnv{\ifmmode {{\rm N_{\rm NV}}} \else ${\rm N_{\rm NV}}$\fi}
\def\nneviii{\ifmmode {{\rm N_{\rm NeVIII}}} \else ${\rm N_{\rm NeVIII}}$\fi}
\def\nmgx{\ifmmode {{\rm N_{\rm MgX}}} \else ${\rm N_{\rm MgX}}$\fi}

\def\vs{\ifmmode {v_s} \else $v_s$\fi}
\def\rs{\ifmmode {r_s} \else $r_s$\fi}

\def\alp{\ifmmode {\alpha} \else $\alpha$\fi}
\def\sig{\ifmmode {\sigma} \else $\sigma$\fi}
\def\tvir{\ifmmode {{\it T}_{\rm vir}} \else ${\it T}_{\rm vir}$\fi}
\def\nvirp{\ifmmode {n_{\rm H}(\rvir)} \else $n_{\rm H}(\rvir)$\fi}
\def\tcgm{\ifmmode {T_{\rm th}(\rcgm)} \else $T_{\rm th}(\rcgm)$\fi}
\def\ncgm{\ifmmode {n_{\rm H}(\rcgm)} \else $n_{\rm H}(\rcgm)$\fi}
\def\alpcgm{\ifmmode {\alpha(\rcgm)} \else $\alpha(\rcgm)$\fi}
\def\mvir{\ifmmode {M_{\rm vir}} \else $M_{\rm vir}$\fi}
\def\rvir{\ifmmode {R_{\rm vir}} \else $R_{\rm vir}$\fi}
\def\vvir{\ifmmode {V_{\rm vir}} \else $V_{\rm vir}$\fi}
\def\xvir{\ifmmode {x_{\rm vir}} \else $x_{\rm vir}$\fi}
\def\xcgm{\ifmmode {x_{\rm cgm}} \else $x_{\rm cgm}$\fi}
\def\rcgm{\ifmmode {R_{\rm CGM}} \else $R_{\rm CGM}$\fi}
\def\mdm{\ifmmode {M_{\rm DM}} \else $M_{\rm DM}$\fi}
\def\mdisk{\ifmmode {M_{\rm disk}} \else $M_{\rm disk}$\fi}
\def\mb{\ifmmode {M_{\rm b}} \else $M_{\rm b}$\fi}
\def\mbm{\ifmmode {M_{\rm b,miss}} \else $M_{\rm b,miss}$\fi}
\def\mbo{\ifmmode {M_{\rm b,obs}} \else $M_{\rm b,obs}$\fi}
\def\fb{\ifmmode {f_{\rm b}} \else $f_{\rm b}$\fi}
\def\mhw{\ifmmode {M_{\rm hot/warm}} \else $M_{\rm hot/warm}$\fi}
\def\mhot{\ifmmode {M_{\rm hot}} \else $M_{\rm hot}$\fi}
\def\mwarm{\ifmmode {M_{\rm warm}} \else $M_{\rm warm}$\fi}
\def\mcool{\ifmmode {\dot{M}_{\rm cool}} \else $\dot{M}_{\rm cool}$\fi}
\def\mheat{\ifmmode {\dot{M}_{\rm heat}} \else $\dot{M}_{\rm heat}$\fi}
\def\mgas{\ifmmode {M_{\rm CGM}} \else $M_{\rm CGM}$\fi}
\def\mstar{\ifmmode {M_{\rm *}} \else $M_{\rm *}$\fi}
\def\mism{\ifmmode {M_{\rm ISM}} \else $M_{\rm ISM}$\fi}
\def\msmbh{\ifmmode {M_{\rm SMBH}} \else $M_{\rm SMBH}$\fi}
\def\mhalo{\ifmmode {M_{\rm halo}} \else $M_{\rm halo}$\fi}
\def\rsun{\ifmmode {R_{\rm 0}} \else $R_{\rm 0}$\fi}
\def\rcool{\ifmmode {R_{\rm cool}} \else $R_{\rm cool}$\fi}

\def\nmean{\ifmmode {\left< n_{\rm CGM} \right>} \else $\left< n_{\rm CGM} \right>$\fi}
\def\zm{\ifmmode {\left< Z' \right>} \else $\left< Z' \right>$\fi}
\def\zmean{\ifmmode {\left< Z'_{\rm CGM} \right>} \else $\left< Z'_{\rm CGM} \right>$\fi}
\def\Tpeak{\ifmmode {T_{\rm peak}} \else $T_{\rm peak}$\fi}
\def\Twarm{\ifmmode {T_{\rm warm}} \else $T_{\rm warm}$\fi}
\def\tgas{\ifmmode {T_{\rm gas}} \else $T_{\rm gas}$\fi}
\def\rgas{\ifmmode {r_{\rm gas}} \else $r_{\rm gas}$\fi}
\def\rogas{\ifmmode {\rho_{\rm gas}} \else $\rho_{\rm gas}$\fi}
\def\mhi{\ifmmode {M_{HI}} \else $M_{HI}$\fi}
\def\nhi{\ifmmode {N_{HI}} \else $N_{HI}$\fi}
\def\rhi{\ifmmode {R_{\rm HI}} \else $R_{\rm HI}$\fi}
\def\rhalf{\ifmmode {r_{1/2}} \else $r_{1/2}$\fi}
\def\thalf{\ifmmode {\theta_{1/2}} \else $\theta_{1/2}$\fi}
\def\tgc{\ifmmode {\theta_{\rm GC}} \else $\theta_{\rm GC}$\fi}
\def\npeak{\ifmmode {N_{\rm peak}} \else $N_{\rm peak}$\fi}
\def\tdyn{\ifmmode {t_{\rm dyn}} \else $t_{\rm dyn}$\fi}
\def\tcool{\ifmmode {t_{\rm cool}} \else $t_{\rm cool}$\fi}
\def\fcool{\ifmmode {f_{\rm cool}} \else $f_{\rm cool}$\fi}
\def\trec{\ifmmode {t_{\rm rec}} \else $t_{\rm rec}$\fi}
\def\Tmin{\ifmmode {T_{\rm min}} \else $T_{\rm min}$\fi}
\def\mbar{\ifmmode {\bar{m}} \else $\bar{m}$\fi}
\def\vturb{\ifmmode {\sigma_{\rm turb}} \else $\sigma_{\rm turb}$\fi}
\def\Psun{\ifmmode {P_{\rm th}(\rsun)} \else $P_{\rm th}(\rsun)$\fi}
\def\s04{\ifmmode {S_{0.4-2.0}} \else $S_{0.4-2.0}$\fi}
\def\Lcool{\ifmmode {L_{\rm cool}} \else $L_{\rm cool}$\fi}
\def\cgm{\ifmmode {\rm CGM} \else ${\rm CGM}$\fi}

\def\nth{\ifmmode {n_{\rm H,thresh}} \else $n_{\rm H,thresh}$\fi}
\def\rth{\ifmmode {r_{\rm thresh}} \else $r_{\rm thresh}$\fi}
\def\dml{\ifmmode {{\rm DM_{\rm LMC}}} \else ${\rm DM_{\rm LMC}}$\fi}
\def\dmt{\ifmmode {{\rm DM_{\rm tot}}} \else ${\rm DM_{\rm tot}}$\fi}

\def\lscale{\ifmmode {{L_s} } \else ${L_s}$\fi}
\def\aL{\ifmmode {a_{\Lambda} } \else $a_{\Lambda}$\fi}
\def\UH{\ifmmode {\left<U\right>_{H}} \else $\left<U\right>_{H}$\fi}
\def\zmin{\ifmmode {\zeta_{min}} \else $\zeta_{min}$\fi}
\def\ypar{\ifmmode {\Tilde{Y}_{500}} \else $\Tilde{Y}_{500}$\fi}
\def\fion{\ifmmode {f_{\rm ion}} \else $f_{\rm ion}$\fi}

\def\kal{\ifmmode {K_{\alpha}} \else $K_{\alpha}$\fi}
\def\kbe{\ifmmode {K_{\beta}} \else $K_{\beta}$\fi}

\def\nh{\ifmmode {n_{\rm H}} \else $n_{\rm H}$\fi}
\def\ca{\ifmmode {\chi_{\rm EM}} \else $\chi_{\rm EM}$\fi}
\def\cb{\ifmmode {\chi_{\rm N}} \else $\chi_{\rm N}$\fi}
\def\lia{\ifmmode {22~{\rm \AA}} \else $22~{\rm \AA}$\fi}
\def\lib{\ifmmode {19~{\rm \AA}} \else $19~{\rm \AA}$\fi}

\def\cg{\ifmmode {c_g} \else $c_g$\fi}
\def\c6{\ifmmode {c_{g,6}} \else $c_{g,6}$\fi}
\def\phim{\ifmmode {P_{\rm HIM}} \else $P_{\rm HIM}$\fi}

\def\chisq{\ifmmode {\chi_{mod}^2} \else $\chi_{mod}^2$\fi}

\def\kb{\ifmmode {k_{\rm B}} \else $k_{\rm B}$\fi}
\def\mp{\ifmmode {m_{\rm p}} \else $m_{\rm p}$\fi}

\def\msun{\ifmmode {\rm M_{\odot}} \else $\rm M_{\odot}$\fi}
\def\lsun{\ifmmode {\rm L_{\odot}} \else $\rm L_{\odot}$\fi}
\def\msuny{\ifmmode {\rm M_{\odot}\:year^{-1}} \else $\rm M_{\odot}\:year^{-1}$\fi}
\def\ergs{\ifmmode {\rm erg\:s^{-1}} \else $\rm erg\:s^{-1}$\fi}

\def\kms{\ifmmode {\rm km\:s^{-1}} \else $\rm km\:s^{-1}$\fi}
\def\cmc{\ifmmode {\rm cm^{-2}} \else $\rm cm^{-2}$\fi}
\def\cmv{\ifmmode {\rm cm^{-3} \:} \else $\rm cm^{-3}$\fi}
\def\kel{\ifmmode {\rm \:\: K} \else $\rm \:\: K$\fi}
\def\pc{\ifmmode {\rm \:\: pc} \else $\rm \:\: pc$\fi}
\def\kpc{\ifmmode {\rm \:\: kpc} \else $\rm \:\: kpc$\fi}
\def\amu{\ifmmode {\rm \:\: amu} \else $\rm \:\: amu$\fi}
\def\fluxun{\ifmmode {\rm \: erg~s^{-1}~cm^{-2}~deg^{-2}} \else $\rm \: erg~s^{-1}~cm^{-2}~deg^{-2}$\fi}
\def\liun{\ifmmode {\rm \:\: photons~s^{-1}~cm^{-2}~sr^{-1}} \else $\rm \:\: photons~s^{-1}~cm^{-2}~sr^{-1}$\fi}

\def\aox{\ifmmode {a_{\rm O}} \else $a_{\rm O}$\fi}

\shorttitle{Exploring the MW CGM with a SAM}
\shortauthors{Faerman et al.}

\begin{document}

\defcitealias{SD93}{SD93}
\defcitealias{Somer08}{S08}
\defcitealias{Somer15}{S15}
\defcitealias{FSM17}{FSM17}
\defcitealias{FSM20}{FSM20}
\defcitealias{Klypin02}{KZS02}

\title{Exploring the Milky Way Circumgalactic Medium in a Cosmological Context with a Semi-Analytic Model}

 \author{
Yakov Faerman \altaffilmark{1 *},
Viraj Pandya \altaffilmark{2,3},
Rachel S. Somerville \altaffilmark{3,4},
and Amiel Sternberg \altaffilmark{5,3,6}
}

\altaffiltext{*}{e-mail: \href{mailto:yakov.faerman@mail.huji.ac.il}{yakov.faerman@mail.huji.ac.il}}
\altaffiltext{1}
{Racah Institute of Physics,The Hebrew University, Jerusalem, Israel}
\altaffiltext{2}
{UCO/Lick Observatory, Department of Astronomy and Astrophysics, University of California, Santa Cruz, CA 95064, USA}
\altaffiltext{3}
{Center for Computational Astrophysics, Flatiron Institute, 162 5th Avenue, New York, NY 10010, USA}
\altaffiltext{4}
{Department of Physics and Astronomy, Rutgers, The State University of New Jersey, 136 Frelinghuysen Road, Piscataway, NJ 08854, USA}
\altaffiltext{5}
{School of Physics and Astronomy, Tel Aviv University, Ramat Aviv 69978, Israel}
\altaffiltext{6}
{Max-Planck-Institut fur Extraterrestrische Physik (MPE), Giessenbachstr., 85748 Garching, FRG}

 \begin{abstract}
 We combine the Santa-Cruz Semi-Analytic Model (SAM) for galaxy formation and evolution with the circumgalactic medium (CGM) model presented in \cite{FSM20} to explore the CGM properties of $L^{*}$ galaxies. We use the SAM to generate a sample of galaxies with halo masses similar to the Milky Way (MW) halo, $\mvir \approx 10^{12}$~\msun, and find that the CGM mass and mean metallicity in the sample are correlated. We use the CGM masses and metallicities of the SAM galaxies as inputs for the FSM20 model, and vary the amount of non-thermal support. The density profiles in our models can be approximated by power-law functions with slopes in the range of $0.75 < a_n < 1.25$, with higher non-thermal pressure resulting in flatter distributions. We explore how the gas pressure, dispersion measure, \ovi-\oviii~column densities, and cooling rates behave with the gas distribution and total mass. We show that for CGM masses below $\sim 3 \times 10^{10}$~\msun, photoionization has a significant effect on the column densities of \ovi~and \oviii. The combination of different MW CGM observations favors models with similar fractions in thermal pressure, magnetic fields/cosmic rays, and turbulent support, and with $\mgas \sim 3-10 \times 10^{10}$~\msun. The MW \ovi~column requires $\tcool/\tdyn \sim 4$, independent of the gas distribution. The AGN jet-driven heating rates in the SAM are enough to offset the CGM cooling, although exact balance is not required in star-forming galaxies. We provide predictions for the columns densities of additional metal ions - \nv, \neviii, and \mgx.
 \end{abstract}


\section{Introduction}
\label{sec_intro}

\citet{Spitzer56} used indirect evidence to infer that the Milky Way disk is embedded in a halo of diffuse warm/hot gas, at $T \sim 10^6$~K. \citet{Bahcall69b} suggested that extended, $\sim 100$~kpc, ``coronae" around other galaxies could explain the absorption features detected in QSO spectra. Analytic calculations and numerical simulations of galaxy formation also predict the existence of hot coronae around massive galaxies \citep{Bregman80b,Cen99,BD03}. This circumgalactic medium (CGM) connects the large scale cosmic web with the galaxies residing at the centers of the dark matter halos. Matter accreting onto halos moves through the circumgalactic volume before it reaches the galaxy, where it can fuel star formation. Galactic feedback, both from stellar processes and from the central supermassive black hole (SMBH), ejects gas and metals into the CGM \citep{Keres09,Nelson14,Marasco15}. This matter may re-accrete back onto the galaxy \citep{Bregman80a,Bertone07,Mari11}, be deposited in the CGM for extended periods of time, or get ejected completely back to the intergalactic medium (IGM), enriching it with metals \citep{Cen06a,Nelson18b}. The CGM influences the evolution of the galaxy and its environment, and is itself shaped by feedback processes taking place within galaxies. It is thus a critical component of galactic systems \citep{TPW17}.

Recent observations reveal a wealth of information about the CGM, showing it is extended, highly multiphase, and contains a significant reservoir of gas and metals \citep{Tumlinson11,Werk14,Peeples14,Werk16,Prochaska17,Burchett19}. However, even for a well-studied galaxy like our own Milky Way (MW), there is still debate about the exact values of the CGM basic properties, such as the gas and metal masses, temperature, and their distributions with radius \citep{Fang13}. These are interesting since they can inform us about the physical processes shaping this diffuse component --- e.g. galactic winds and outflows \citep{Sarkar15,Fielding17a,Li20a,Schneider20}, thermal instabilities \citep{McCourt18,Liang20}, metal diffusion and mixing, magnetic fields \citep{Sparre20,Voort21} and cosmic rays \citep{Butsky20,Ji20}. 

\citet{Pandya20} have shown that the CGM provides a powerful means of discriminating between and constraining feedback processes, so the rapidly accumulating archive of observations is a critical resource that should be exploited to better constrain theoretical models. However, predicting observables from currently available models and simulations is not straightforward. The CGM has been studied in numerical cosmological hydrodynamic simulations, and here also shows great thermodynamic and kinematic complexity \citep{Nelson16,Gutcke17,Oppen18b,Hafen19a,Fielding20b}. The multi-scale physical processes affecting galaxy formation require a high dynamic range, resulting in high computational costs and limited spatial resolution in the CGM \citep{Peeples19,Hummels19,Voort19}. Small-scale processes, such as star formation and stellar and AGN-driven feedback, are often implemented using sub-grid models. However, uncertainties in our understanding of these processes and their coupling to the large scale structure and evolution of the CGM add to the challenge of numerical experiments \citep[for reviews see][]{SD15,naab:2017}. These numerical techniques are typically too expensive to carry out a systematic study of a cosmologically representative sample of many halos, or to explore the implications of adopting different physical ingredients or sub-grid recipes. 

Semi-analytic models provide another powerful tool to study galaxy formation in a cosmological context \citep[e.g.][]{Kauffmann93,Somer99,Cole:1994}. For a review of more recent work with semi-analytic models, see \citet{SD15}.  These are set within cosmological "merger trees", which describe the formation histories of the underlying dark matter halos, and use a set of coupled ordinary differential equations and empirical prescriptions to self-consistently describe the key processes governing galaxy evolution, including accretion of gas onto galaxies, star formation, metal enrichment, and feedback. Another advantage of these models is their low computational cost, allowing flexibility by trying different parameter values and physics prescriptions. Up to now, SAMs have been calibrated and tested using mainly UV-optical observations that probe the stellar mass content and recent star formation history of galaxies \citep[e.g.][]{Somer08,Somer15,Somerville:2021} or sub-mm/radio observations that probe the cold ISM content \citep{popping:2014,popping:2019}. Predictions for the CGM are usually not even presented, and to our knowledge there is currently no published work that attempts to make a direct comparison between SAM predictions and CGM observations. 

Analytic models have been employed to study and make predictions for the detailed properties of the CGM, with many recent works exploring different physical assumptions and gas distributions \citep{MB04, Anderson10, Miller13, FSM17, MP17, MW18, Stern18, Qu18b, Voit19,FSM20}.  These models can be compared with both observations and numerical simulations, providing a better understanding of the main mechanisms shaping the CGM and galaxies in general. Analytic models can also be used to fit measurements of individual galaxies and infer quantities that are not measured directly, such as gas mass, temperature, etc. 

\citet[hereafter FSM20]{FSM20} presented a detailed model for the CGM of MW-mass galaxies in the low-redshift Universe. Their fiducial set of model parameters was successful in reproducing the X-ray absorption observations of the MW \citep{Bregman07,Gupta12,Fang15} and the \ovi~measurements from the COS-Halos and eCGM surveys \citep{Tumlinson11,Werk14,Johnson15}. \citetalias{FSM20} did not present results for how the CGM observables are affected by variations in the model parameters given the possible dispersion in CGM properties across different MW-like galaxies.

In this work we combine the FSM20 model with the Santa Cruz SAM \citep{Somer08,Somer15} to explore the CGM of MW-mass galaxies. We use the CGM mass and metallicity calculated by the SAM for a suite of realizations of MW-mass halos as inputs for the FSM20 model. Our goals are to (i) show how variations in the FSM20 fiducial model input parameters affect the CGM predictions (ii) compare the predictions of CGM observables with MW observations to test the models (iii) build a framework that we plan to use in future works to make predictions of CGM observables for a broader suite of halo masses and redshifts.

This paper is structured as follows: in Section \ref{sec_sams} we briefly summarize the Santa Cruz SAM framework and present our galaxy sample. We recap the FSM20 CGM model in Section~\ref{sec_fsm}, discuss the model variations we consider in this work, and summarize the observational constraints from the MW CGM in Section~\ref{sec_mw_obs}. We then present the results of our modeling in Section~\ref{sec_res}. We show the gas distributions, examine the behaviour of oxygen columns and compare them to MW observations, and address the gas cooling rates. We discuss our results in Section~\ref{sec_dis} and summarize in Section~\ref{sec_summ}.

\section{The Santa Cruz Semi-Analytic Model}
\label{sec_sams}

In this section we describe the galaxy formation model we use in this work. We briefly present the SAM and its treatment of the CGM (\S~\ref{subsec_sam_frame}), describe how we use the SAM to construct a sample of galaxies (\S~\ref{subsec_sam_samp}), and discuss their properties (\S~\ref{subsec_sam_prop}).

\subsection{SAM Framework}
\label{subsec_sam_frame}

The Santa Cruz SAM is a framework for modeling galaxy formation and evolution (\citealp{Somer99, Somer01, Somer08, Somer15, Somerville:2021}). Using analytic prescriptions, the model follows the growth of dark matter halos and the galaxies that form in them in a $\Lambda$CDM~cosmology, tracing their stellar populations, super-massive black holes (SMBHs), metal enrichment, and the inflows and outflows of diffuse gas. The model includes feedback from active galactic nuclei (AGN), in the form of winds that eject cool gas from the galaxies and energy injection that heats gas around galaxies and in galaxy clusters. The SAM successfully reproduces many different observations of galaxies in the Universe, such as the galaxy mass and luminosity functions, star formation rates and their relation to galaxy masses, and the correlation between the SMBH and galactic bulge mass \citep[][hereafter S08]{Somer08}. A recent update included the detailed treatment of the multiphase interstellar medium \citep[][hereafter S15]{Somer15}. 

Fundamentally, the SAM tracks the flows of gas between different reservoirs, such as the pristine intergalactic medium (IGM), the hot halo gas, the cold interstellar medium (ISM), and the stellar disk and bulge. Gas that has been heated by supernovae feedback and ejected from the ISM is deposited in one of two reservoirs: the hot halo, assumed to be at a constant temperature \tvir, and residing within \rvir, or an ``ejected" reservoir, which re-accretes into the halo and also becomes available for cooling but on a longer timescale. The actual physical state and spatial location of the ``ejected" reservoir is ambiguous: it can be conceived of either as gas that is physically ejected and resides outside the halo, possibly heated to temperatures greater than \tvir, or as gas that remains within the halo but cannot cool efficiently. In our fiducial models, we associate the classical SAM ``hot halo" reservoir with the CGM. We discuss models for which the CGM is assumed to be the sum of the two components (hot halo and ejected) in \S~\ref{subsec_dis_mass}.

\subsection{SAM Calibration and Galaxy Sample Selection}
\label{subsec_sam_samp}

We generate a catalog of galaxies by running the Santa Cruz SAM on four sub-volumes of the {\it Bolshoi-Planck} simulation, each with a box size of $50^3~{\rm Mpc}^3$. The SAM assumes the \citet{Planck16} cosmology, with $h = 0.678$, $\Omega_{m,0} = 0.308$, $\Omega_{\Lambda,0}= 0.692$, and $\Omega_{b,0} = 0.0486$. These are compatible with the \citet{Planck14} cosmology assumed in the Bolshoi-Planck simulation (see \citealp{RP16} for more details). The halo catalogs and merger trees we use are based on Rockstar \citep{Behroozi13a} and consistent-trees \citep{Behroozi13b}, respectively.
The astrophysical parameters are specified in \citetalias{Somer15}, with some parameter values recalibrated to update the models to the Planck cosmology. The observations used for the calibration and the results of the calibration comparison are shown in \citet{Yung:2019a} Appendix~B. The calibration quantities include the stellar mass function, the stellar mass vs.~cold gas fraction, stellar mass vs.~metallicity relation, and the bulge mass vs.~SMBH mass relation. The updated parameter values for the Bolshoi-Planck cosmology are specified in \citet{Somerville:2021}.

We focus on MW-mass halos, and select from the catalog objects with halo masses in the range $\mvir = 0.8 - 1.2 \times 10^{12}$~\msun~at $z=0$~(see \citealt{PH19}). As we show in the next section, the baryonic mass of these galaxies is dominated by the stellar and CGM mass. Due to numerical artifacts, some halos have baryonic fractions greater than the cosmological fraction, given by $f_{\rm bar} = \Omega_{b,0}/\Omega_{m,0} \approx 0.158$. We remove these halos, which constitute $5\%$ of the total halos in this mass range, from our sample.
Some SAM galaxies have very low CGM masses, implying mean CGM densities that are only barely above the mean density of baryons in the Universe. For the cosmological parameters used in the SAM, $\left< \rho_{b} \right>_{\rm cosm} \approx 4.2 \times 10^{-31}$~gr~\cmv, and within the extent of the circumgalactic medium in our CGM model ($\rcgm = 283$~kpc, see \S~\ref{sec_fsm}) this is equal to a gas mass of $5.9 \times 10^{8}$~\msun. We discard objects with $\mgas < 6 \times 10^{9}$~\msun, with mean over-densities of $\lesssim 10$. For a MW-mass halo, this translates to $5\%$ of the galactic baryonic budget in the CGM. Lower gas masses or baryonic fractions may not be impossible but are somewhat extreme in massive dark matter halos, and inconsistent with the extended CGM assumed in \citetalias{FSM20}.

Our final sample includes galaxies that obey these three constraints --- MW-like halo mass, maximum baryonic fraction of the cosmological mean, and minimum baryonic mean density. While the SAM can generate as many of these as needed, in our analysis we show $\sim 600$ objects, sampling a wide range of CGM masses and metallicities, and we now present and discuss their properties.

\subsection{SAM Galaxy Sample Properties}
\label{subsec_sam_prop}

 \begin{figure*}[t]
 \includegraphics[width=0.99\textwidth]{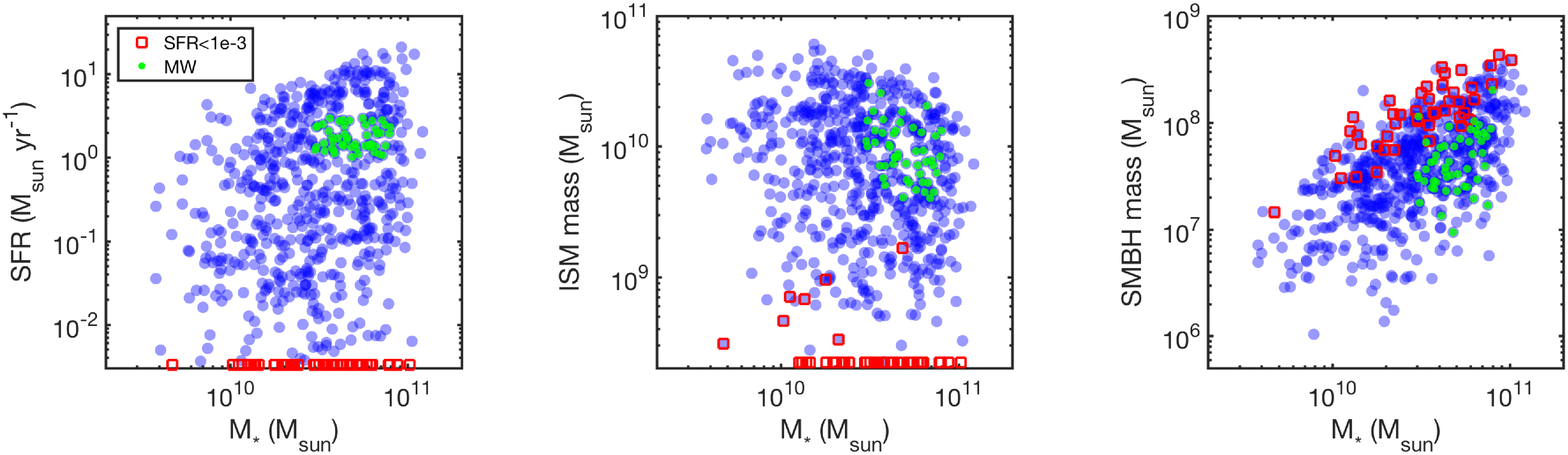}
\caption{Properties of the SAM-generated galaxies as functions of their stellar mass (\mstar) --- the (instantaneous) star formation rate (SFR, left), mass of the interstellar medium (\mism, middle), and the supermassive black hole (\msmbh, right). The red markers show objects with ${\rm SFR} < 10^{-3}$~\msuny, outside the plotted SFR range. The SFR and \mism~show no correlation with \mstar~over the limited halo mass range in our sample. At low stellar masses, \mism~can be higher than \mstar~but it never dominates the baryonic mass in the halo (see Figure~\ref{fig:galcgm}). The \msmbh~in the SAM is correlated with the bulge mass, and shows some correlation with the total \mstar. The green circles show a subsample with stellar mass and SFR similar to the MW~(see \S~\ref{subsec_sam_prop} for details).}
  \label{fig:galprop}
\end{figure*}

\begin{figure*}[t]
 \includegraphics[width=0.99\textwidth]{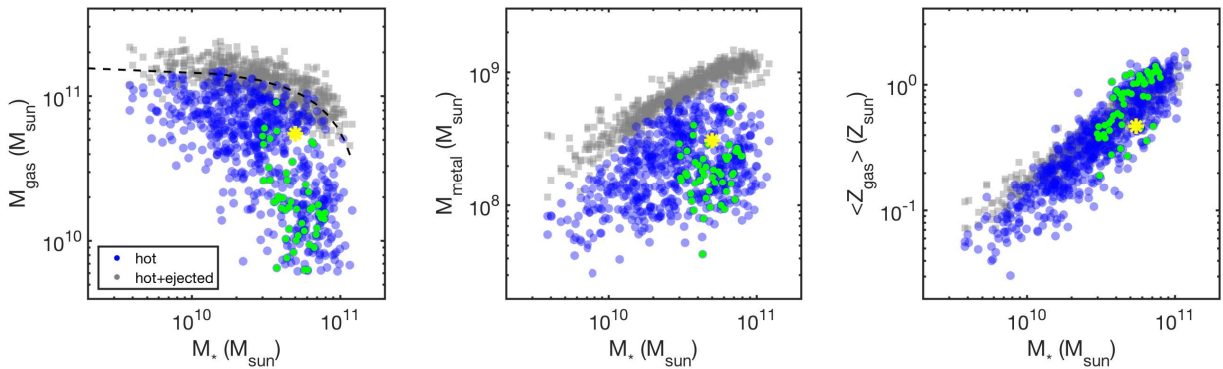}
\caption{The extended gaseous baryons in the SAM galaxies, with the hot halo component (blue circles), and the total diffuse gas mass (hot halo+ejected, grey squares). The left and middle panels show the mass of gas and metals, respectively, as functions of the galaxy stellar mass. The dashed curve in the left panel shows the remaining cosmological baryonic mass of a $10^{12}$~\msun~halo after subtraction of $M_*$, and the blue markers are below this curve as expected. The right panel shows the mean (mass-averaged) gas metallicity. The green circles show the MW-like galaxies, and these have a range of CGM masses and metallicities similar to the full sample. The yellow markers show the values of the FSM20 observationally-motivated fiducial model (see \S~\ref{subsec_sam_prop} for details).}
  \label{fig:galcgm}
\end{figure*}

Figure~\ref{fig:galprop} shows several properties of the galaxies in our sample as functions of the stellar mass, \mstar. The left panel shows the instantaneous star formation rate (SFR)\footnote{~The mean SFRs, averaged over timescales of a few hundred Myr to $1$~Gyr, are similar to the instantaneous rates and we do not show them.}. For $90\%$ of the galaxies, the SFR is between $0.01$ and $10$~\msuny, and the red squares show objects with ${\rm SFR} < 10^{-3}$~\msuny, outside the plotted range. The middle panel shows the mass of the interstellar medium (ISM). For galaxies with low stellar masses, the ISM mass can be higher than \mstar, but as we show in Figure~\ref{fig:galcgm}, the CGM component of these objects is even more massive and the ISM never dominates the galactic baryon budget. The mass of the SMBH in the SAM (right panel) is correlated with the bulge mass as enforced by the BH growth model in the SAM \citepalias{Somer08} and weakly correlated with the total stellar mass of the galaxy. Objects with the lowest SFR (red squares) have the highest SMBH masses at their stellar mass, which is a direct result of AGN feedback quenching star formation. Versions of these diagrams for a wider halo mass range, for a cosmologically representative sample, are shown in \citetalias{Somer15} and \citetalias{Somer08}. 
In \S~\ref{subsec_dis_ebudg} we discuss the energy budget of the CGM and the heating rate by the AGN. 

The green circles in Figure~\ref{fig:galprop} mark a sub-sample of galaxies similar to the MW in stellar mass and SFR. We select objects with $M_* = 3-8 \times 10^{10}$~\msun~\citep{BHG16}, and ${\rm SFR}=1-3$~\msuny~\citep{Lic15}, resulting in a sample of $\sim 50$ galaxies. We note that the median SMBH mass of this sample is $\approx 4 \times 10^{7}$~\msun, more massive than the MW SMBH, with $\approx 4.1 \times 10^{6}$~\msun~\citep{Boehle16,Grav19}. However, the MW SMBH mass is unusual in the full sample, and adding it as a selection criterion would leave only a small number of objects. We show the MW-like subsample in the following figures in this work to allow comparison to MW CGM properties and observables.

Figure~\ref{fig:galcgm} shows the properties of the extended gaseous components in our SAM galaxies, as functions of the stellar mass. The blue circles show the hot halo, and the grey squares show the sum of the hot halo and ejected components (see \S~\ref{subsec_sam_frame}). In our fiducial models, we equate the hot halo component with the CGM.

The left and middle panels present the masses of gas and metals, respectively, as functions of \mstar. The range of the total gas masses (CGM+ejected) is small, with $90\%$ of the objects in the range between $0.66$ and $1.8 \times 10^{11}$~\msun. This is a result of the SAM assumption that nearly all of the baryons accreted into the halo are retained either in the CGM or the ejected reservoir, and only gas ejected by the radiatively efficient mode of AGN feedback is assumed to be removed completely. The black dashed line shows the prediction $M_{\rm gas} = f_b M_{\rm halo}-\mstar$~for $M_{\rm halo}=10^{12}$~\msun, and the total gas masses lie along this line, with some scatter due to the range of halo mass in our sample. The CGM mass also decreases with increasing stellar mass, although more steeply, with a $90\%$ range of $0.85-10.8 \times 10^{10}$~\msun. At low stellar masses, the diffuse baryons around the galaxy are dominated by the CGM and the ejected gas mass is low. For high \mstar, the diffuse gas is dominated by the ejected component, resulting from significant stellar feedback. The SAM version we use in this work does not include heating of the CGM by stellar feedback, and in this exploratory study, we do not link the SFR to the CGM properties.

The mass of metals in the diffuse components (middle panel) increases with the stellar mass, as expected. The correlation is tighter when looking at the total metal mass (CGM+ejected), similar to the behaviour of the gas mass in the left panel. The right panel shows the mean, mass-weighted, metallicity of the diffuse components relative to solar, \zm. Both the CGM and the total gas mean metallicities are strongly correlated with the stellar mass of the galaxy and follow a linear relation with a slope of unity. Most ($\sim 90\%$) of the objects have CGM metallicities between $0.1$ and $1.1$ solar.

The CGM gas and metal masses of the MW-like sample (green circles) are similar to those of the full sample, at a given stellar mass. The yellow marker in Figure~\ref{fig:galcgm} shows the values of the observationally-based FSM20 fiducial model, with $\mgas = 5.5 \times 10^{10}$~\msun, $M_{\rm metals} = 3.1 \times 10^{8}$~\msun, and $\zmean \approx 0.5$. The SAM hot halo properties are consistent with \citetalias{FSM20}, without tuning.

 \begin{figure}[t]
 \includegraphics[width=0.48\textwidth]{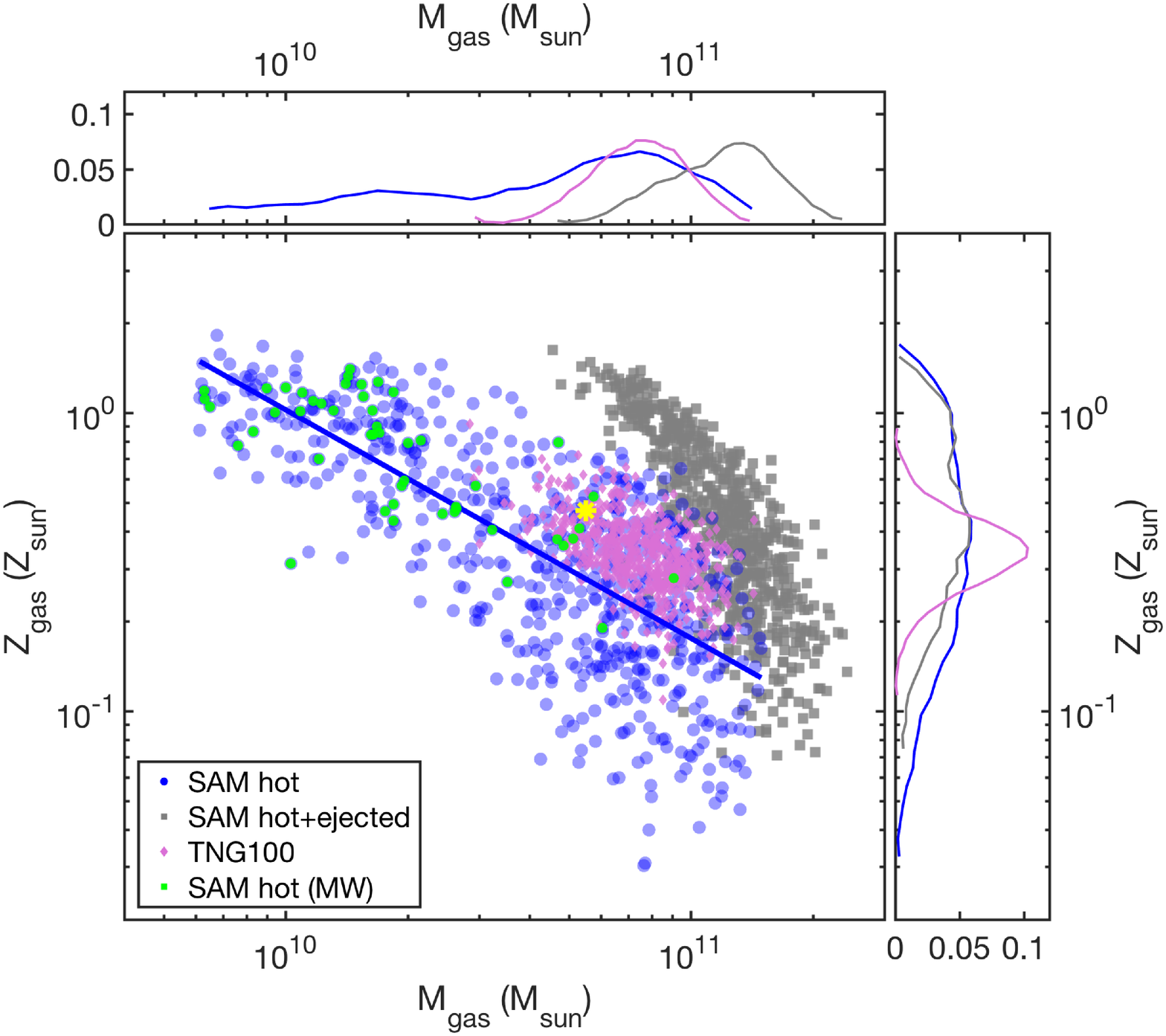}
\caption{The mass-metallicity, or $M-\zm$, relation for the CGM (see \S~\ref{subsec_sam_prop} for details). The blue markers show the hot CGM in the SAM, which roughly follow a power-law function with a slope of $a \approx -0.75$, shown by the thick blue curve. For comparison, galaxies in the Illustris-TNG100 simulation (magenta points) exhibit a similar slope, with narrower distributions. The grey markers show the total extended gas (CGM+ejected) in the SAM, for which the metallicity is a strong function of the gas mass. The green circles show a subsample with stellar mass and SFR similar to the MW. The panels on the top and right show histograms of the gas mass and metallicity, respectively, for each of the samples.}
  \label{fig:galcgm2}
\end{figure}

Figure~\ref{fig:galcgm2} plots the mean metallicity of the extended components, versus the gas mass, $M_{\rm gas}$. We refer to this as the CGM mass-metallicity, or $M-\zm$, relation. For the CGM, the mean metallicities range from $0.03$ to $1.8$ solar, and decrease with CGM mass. This relation can be fitted with a power-law function
\beq \label{eq:mzrel}
\zmean \approx 0.3 \times \left( \mgas/ 5 \times 10^{10}~\msun \right)^{-0.75} ~~~,
\eeq
shown in the figure by the solid blue line. There is a significant scatter at a given gas mass, and we use the fit only for qualitative discussion. The metallicity of the total gas mass varies by over an order of magnitude for a small range in gas mass ($\pm 0.3$~dex), and is a very strong function of the gas mass. The MW-like sample has CGM masses between $0.63$ and $9.1 \times 10^{10}$~\msun, and mean metallicities between $0.20$ and $1.4$ solar.

We compare the $M-\zm$ relation in the SAM to that of galaxies from the TNG100 simulation \citep{Weinberger17,Pillepich18,Nelson18a}. We select galaxies with halo masses in the range $0.8-1.2 \times 10^{12}$~\msun~from the simulation halo catalog. We define the CGM as cells between $0.1$ and $1.0$~\rvir~with ${\rm SFR}=0$, and use the particle data to measure the CGM masses and mean metallicities (see Cohen et. al., in prep.). The results are plotted by the magenta circles in Figure~\ref{fig:galcgm2}, and are overall similar to the CGM $M-\zm$ relation in the SAM, though with a smaller scatter. The median metallicity in the TNG100 sample is $0.36$, close to the value in the SAM sample, and the median CGM mass is $7.4 \times 10^{10}$~\msun, higher by a factor of $1.4$ than in the SAM. Both the SAM and the TNG hot CGM properties are consistent with the FSM20 fiducial model (yellow marker) without tuning.

\section{The FSM20 CGM model}
\label{sec_fsm}

\citetalias{FSM20} presented a phenomenological model for the CGM of MW-mass galaxies, reproducing UV and X-ray observations of the MW and other star-forming galaxies in the low-redshift Universe. In this section we summarize the FSM20 framework (\S~\ref{subsec_fsm_prop}) and present the model parameter space we explore in this work (\S~\ref{subsec_fsm_space}).

\subsection{Model Framework}
\label{subsec_fsm_prop}

The FSM20 model describes a large-scale, spherically symmetric corona, with gas in hydrostatic equilibrium (HSE) in the gravitational potential of a MW-mass dark matter halo. The model adopts a barotropic equation of state (EoS), assumes constant entropy, and allows for different pressure components, including thermal pressure, magnetic fields (B), cosmic rays (CR), and turbulent support. The gas ionization states and cooling efficiencies are set by collisional ionization and photoionization by the metagalactic radiation field (MGRF).

A radial metallicity profile is included, and is given by $Z'(r)  = Z'_{in} (1+x^2)^{-1/2}$, where $x = r/r_Z$, and $r_Z$ is the metallicity length scale. This two-parameter function can be defined either by the gas metallicity at one boundary and $r_Z$, or the gas metallicities at the inner and outer boundaries, $Z'_{\rm in}$ and $Z'_{\rm out}$, respectively. The individual elemental abundances are taken from \cite{Asplund09}.

In this work we fix some of the parameters to the values adopted in the FSM20 fiducial model, and vary others to examine their effect on the CGM properties. Table~\ref{tab:mod_prop} summarizes the model input parameters and the values used here. As described in \S~\ref{subsec_sam_samp}, we focus on dark matter halos similar to that of the Milky Way, with $\mvir \approx 10^{12}$~\msun. The virial radius for this mass in \citetalias{Klypin02} and the SAM is $\rvir\approx 260$~kpc. \citetalias{FSM20} extend the CGM to $\rcgm=1.1 \rvir \approx 283$~kpc, motivated by the \ovi~measurements from the eCGM survey \citep{Johnson15}, and we adopt this value. In \citetalias{FSM20}, the gas temperature at the outer boundary is set to the temperature of the virial shock (see Equations (13)-(14) there), and we fix it with $\tcgm = 2.4 \times 10^5$~K in the FSM20 fiducial model. The inner boundary of the CGM is taken to be the solar radius, $\rsun = 8.5$~kpc.

We examine two variations. First, for a given density and metallicity profile shape, we vary the total CGM mass (or mean gas density) and its mean metallicity. We change~\mgas~by adjusting the gas density at the outer boundary, \ncgm~and the mean gas metallicity is controlled by $Z'_{\rm out}$. Second, we vary the shapes of the distribution of metals, by changing the metallicity length scale, $r_Z$, and the gas distribution by adjusting the amount of non-thermal support, which we discuss next.

 \bgroup
 \def\arraystretch{1.2}
 \begin{table}
 \centering
 	\caption{Model Properties}
 	\label{tab:mod_prop}
 		\begin{tabular}{| l || c | c | c |}
 			\midrule
 			\multicolumn{4}{| c |}{Input Parameters - Fixed  (see \S~\ref{subsec_fsm_prop})}	\\
 			\midrule
  			$\mvir$ 				    & 	\multicolumn{3}{| c |}{$10^{12}$~\msun}       \\
 			$\rvir$ 				    & 	\multicolumn{3}{| c |}{$258$~kpc}             \\
 			$\rsun$ 				    & 	\multicolumn{3}{| c |}{$8.5$~kpc}  \\
 			$\rcgm$ 				    & 	\multicolumn{3}{| c |}{$1.1\rvir = 283$~kpc}  \\
 			$\tcgm$ 				    & 	\multicolumn{3}{| c |}{$2.4 \times 10^{5}$~K} \\
 			\midrule
 			\multicolumn{4}{| c |}{Input Parameters - Varied (see \S~\ref{subsec_fsm_space})}	\\
 			\midrule
 			                            & thermal & standard & non-thermal \\
 			\midrule
 			$\vturb~(\kms)$             & $1$	& $60$  & $100$     \\
 			$\alp(\rcgm)$               & $1.1$	& $2.1$ & $2.9$     \\
 			$\alpha_{\rm tot}(\rcgm)$   & $1.1$	& $3.2$ & $5.9$     \\
 			
 		    $r_Z$ (kpc)					& $30$	& $90$  & $250$     \\
 		    $Z'_{\rm in}/Z'_{\rm out}$	& $10$	& $3.3$ & $1.5$     \\
            $n_{\rm H}(\rcgm)$ ($10^{-5}$~\cmv)	& $0.08-1.7$	& $0.13-2.6$  & $0.15-3.1$ \\
 			$Z'_{\rm out}$ (solar)		& $0.05-0.8$ & $0.06-1.1$ & $0.08-1.5$ \\
            \midrule
 			\multicolumn{4}{| c |}{Key Output Properties} \\
 			\midrule
 			$a_n$                       &	$1.2$  & 	$1.0$   & $0.8$    \\
 			$\gamma_{eff}$              &	$1.66$ &  	$1.37$  & $1.25$    \\
 			$T(\rsun)$ ($10^6$~K)       & 	$3.1$  &    $2.4$	& $1.4$    \\
 			
 			\bottomrule
 	\end{tabular}
 \end{table}

\subsection{Model Parameter Space}
\label{subsec_fsm_space}

\begin{figure*}[t]
 \includegraphics[width=0.99\textwidth]{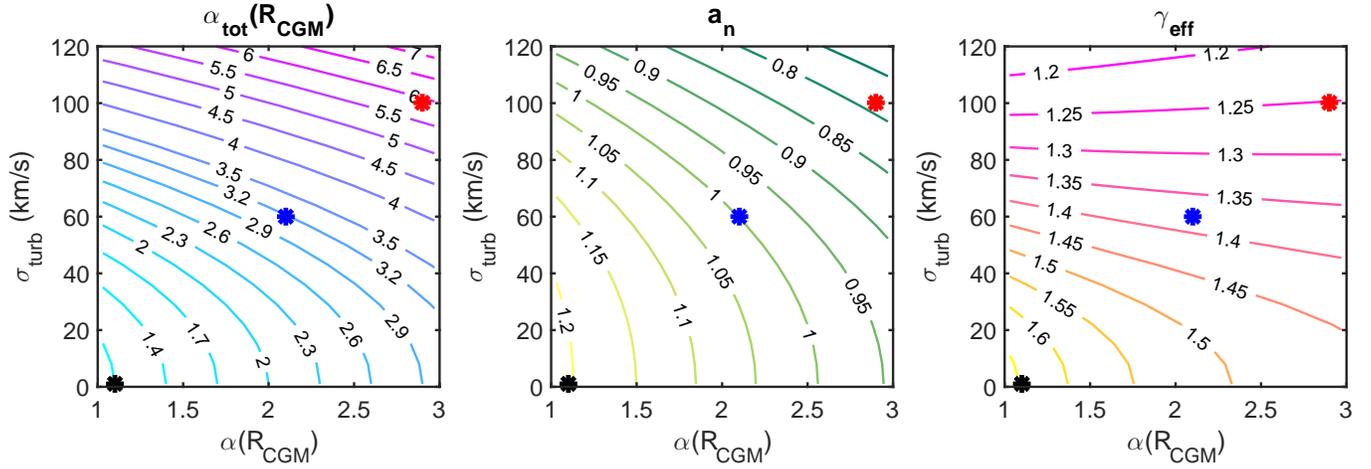}
 \caption{The gas distribution properties in the FSM20 models as functions of the non-thermal support parameters, $\alpcgm$ and $\vturb$ (see \S~\ref{subsec_fsm_space} for details). {\bf Left:} The total to thermal pressure ratio, $\alpha_{\rm tot}$ at \rcgm~(see Eq.~\eqref{eq:atot}). {\bf Middle:} The slope of the power-law approximation of the gas density radial distribution, $a_n$. {\bf Right:} The effective adiabatic index, relating the gas total pressure to the density (see Eq.~\eqref{eq:geff}). The markers in the panels show the three parameter sets we examine in this work (see Table~\ref{tab:mod_prop}) - thermal (black), standard (blue) and non-thermal (red).}
  \label{fig:fsm_space}
\end{figure*}

The shapes of the gas density and pressure profiles depend on the FSM20 input parameters. For a fixed halo potential and temperature \tcgm, the hydrostatic gas density distribution is set by two parameters. First is the ratio of B/CR to thermal pressure, $P_{\rm BCR}/P_{\rm th} \equiv \alp-1$. The two components have different adiabatic indices, so that \alp~is a function of radius. We define \alp(\rcgm)~as the value at the outer CGM boundary. Second is the turbulent velocity scale, \vturb, which we assume is constant throughout the halo. We consider $1<\alpcgm<3$ and $0<\vturb<120~\kms$.

The ratio of total to thermal pressure at any radius is given by
\begin{equation}\label{eq:atot}
\alpha_{\rm tot} =  \frac{P_{\rm tot}}{P_{\rm th}} = \alpha + \frac{\mbar \sigma_{\rm turb}^2}{T_{\rm th}} = \alpha + \mathcal{M}^2 ~~~.
\end{equation}
where $\mathcal{M} = \vturb/c_s$ is the Mach number of turbulence. The gas temperature is a function of radius and $\vturb=const.$, leading to $\mathcal{M}$ and $\alp_{\rm tot}$ varying with radius. The left panel in Figure~\ref{fig:fsm_space} shows $\alpha_{\rm tot}$ at \rcgm. For the parameter ranges we explore, $\alpha_{\rm tot}(\rcgm)$ varies from $1$ to $\sim 7$.

As we show in \S~\ref{sec_res}, the gas density profile in our models can be well-approximated by a power-law function, $n \propto r^{-a_n}$, and the middle panel of Figure~\ref{fig:fsm_space} shows the values of $a_n$. For $\alpcgm \sim 1$~and $\vturb \sim 0$, support is provided by thermal pressure, and $a_n \sim 1.25$. The profile flattens as non-thermal support increases, either by increasing the B/CR pressure, the turbulence, or both. For models dominated by non-thermal pressure, with $\alpcgm \sim 3$ and $\vturb \sim 120~\kms$, $a_n \sim 0.75$. The overall range of $a_n$ is relatively narrow, and centered at unity\footnote{~The slope we use here is obtained using a linear spacing in radius, whereas in \citetalias{FSM20} we used logarithmic spacing. The linear (logarithmic) spacing produces more accurate fits at large (small) radii, close to \rcgm~(\rsun). The two methods give identical values of $a_n$ for models dominated by non-thermal support. However, for profiles in which thermal support is more dominant, logarithmic spacing results in steeper slopes, with $a_n \sim 1.03$ for models with only thermal support.}. The accuracy of the power law approximation is high for models with high non-thermal support, with an error of $<5\%$ at all radii between \rsun~and~\rcgm~for density distributions with $a_n \sim 0.8$. For models dominated by thermal support the approximation is less accurate, with and error of $\lesssim 40\%$ at radii above $\sim 20$~kpc for models with $a_n \sim 1.2$.

As a result of the barotropic EoS, the gas temperature is related to the density, and the thermal temperature profile can be also approximated as $T_{\rm th} \propto r^{-a_T}$, with $a_T = 2a_n/3$. For a given temperature at the outer boundary, the temperature in the inner part will be higher (lower) for steeper (flatter) density profiles. In our models, the temperature at $\rsun$~varies between $1.4 \times 10^6$~K and $3.1 \times 10^6$~K.

The different pressure mechanisms behave differently with density and radius. We define the effective adiabatic index as
\begin{equation}\label{eq:geff}
\gamma_{\rm eff} \equiv \left< \frac{dln(P_{\rm tot})}{dln(n)} \right> ~~~, 
\end{equation}
where the averaging is volume-weighted. This allows us to write an effective equation of state for the CGM. The total pressure profile can then be approximated by $P_{\rm tot} \propto r^{-a_n \gamma_{\rm eff}}$, and $\gamma_{\rm eff}$ is plotted in the right panel of Figure~\ref{fig:fsm_space}, varying from $5/3$ for thermal models to $\approx 1.2$ for models with significant turbulent and B/CR support.

The markers in Figure~\ref{fig:fsm_space} show the parameter sets we consider, spanning the parameter space of \alpcgm~and \vturb. The blue marker shows the standard model, with parameters identical to the FSM fiducial model, in which the three components contribute similarly to the total pressure at \rcgm. The black and red markers show models dominated by thermal and non-thermal pressure, respectively. The parameters for all three models are given in Table~\ref{tab:mod_prop}, and we describe these in more detail in \S~\ref{sec_res}.

\begin{figure}[t]
\centering
\includegraphics[width=0.43\textwidth]{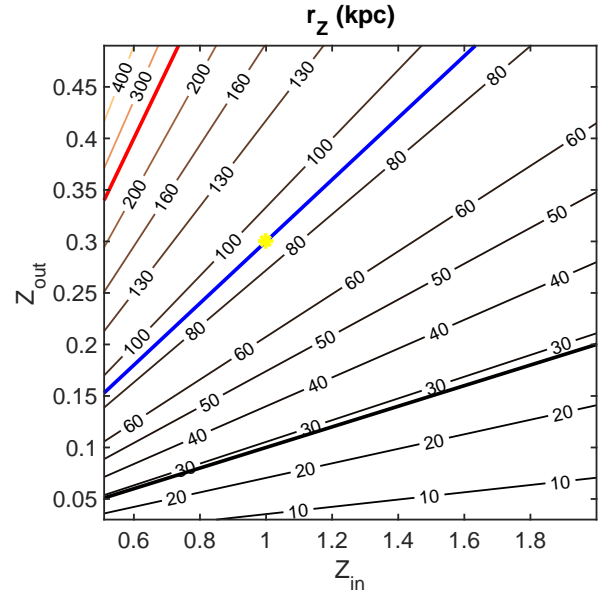}
 \caption{The metallicity profile length scale, $r_Z$, as a function of the metallicities at the CGM boundaries, \rsun~and \rcgm~(see \S~\ref{subsec_fsm_space}). The thick contours show the length scales we adopt for the gas distributions shown in Figure~\ref{fig:fsm_space}. The yellow marker indicates the metallicity values used in \citetalias{FSM20}.}
  \label{fig:fsm_met}
\end{figure}

The shape of the gas metallicity distribution can be defined either by the ratio of the metallicities at the boundaries, or by the metallicity length scale, $r_Z$. Figure~\ref{fig:fsm_met} links these two parameter combinations and shows $r_Z$ as a function of $Z'_{\rm out}$ and $Z'_{\rm in}$, the metallicities at \rcgm~and \rsun, respectively. The length scale changes from $\sim 10$~kpc for a large metallicity gradient, $Z'_{\rm in}/Z'_{\rm out}>20$, to $r_Z > \rcgm$ for metallicity profiles that are close to constant, with $Z'_{\rm in}/Z'_{\rm out}<1.3$. The thick contours mark the metallicity scale lengths we use in this work, with the black (red) curve showing the steep (flat) metallicity distributions. The thick blue contour shows the scale length adopted in the standard model, and the yellow marker shows the specific boundary metallicity values adopted in \citetalias{FSM20}.

Next, for each parameter combination defined by the gas distribution shape (standard, thermal and non-thermal), we construct a series of profiles with the gas masses and mean metallicities given by the SAM. We want to examine how the CGM physical properties and observables calculated by the FSM20 model vary as a function of gas mass and profile shape. We also compare the different observables to the MW measurements, aiming to constrain the MW CGM distribution and mass. To do this, we now revisit and summarize the CGM observations available for the MW.

\section{Milky Way observations}
\label{sec_mw_obs}

\citet[hereafter FSM17]{FSM17} summarized several MW CGM observations relevant to our work (see their Section~2 and Table~1). We adopt their values for the \ovii~and \oviii~column densities, and the \ovii/\oviii~ratio. The nominal values and the $1-\sigma$ errors are given in Table~\ref{tab:obs_prop}. We also adopt their ranges for the thermal pressure above the Galactic disk, inferred from observations of High Velocity Clouds (HVCs).

For the dispersion measure, we consider two quantities. First, the DM to the Large Magellanic Cloud (LMC), at $d \sim 50$~kpc, has been measured using observations of pulsars in the LMC. \cite{Anderson10} provide an upper limit for the MW CGM contribution, with $<23$~pc~\cmv, and \cite{PZ19} estimate $\dml = 23 \pm 10$~pc~\cmv. Second, the total DM in the CGM can be measured using Fast Radio Bursts (FRBs) at cosmological distances~\citep{Thor13,Prochaska19}. This measurement also includes contributions from the FRB host galaxy and the IGM. A large sample, of thousands of localized FRBs with measured redshifts, will allow the contribution of the IGM to be subtracted, minimize the uncertainty from the host, and provide a strong constraint on the DM of the MW CGM. Such a sample is not available yet. However, \cite{Platts20} recently demonstrated a new method of inferring the MW CGM contribution to the DM of FRBs. Applying it to existing measurements they report ${\rm min[\dmt]} = 63^{+27}_{-21} \pm 9$~pc~\cmv, where the first error is statistical and the second - systematic. They also estimate a conservative $95\%$ limit of $\dmt < 123$~pc~\cmv~and warn that the current data set does not allow to rule out or differentiate between existing CGM models.

We also consider the FUSE \ovi~observations analyzed by \citet[hereafter S03]{Savage03}. They report column density  measurements for 84 sightlines, with a mean (median) column and dispersion of ${\rm log(\novi)}= 13.95 ~(13.97) \pm 0.34$. The mean and median are very close and we adopt the median for the nominal value. In their measurements, S03 apply a velocity cut to their spectra, to avoid contamination from gas in the Galactic disk. \cite{Zheng15} show that \ovi~``hidden" in the low-velocity CGM may have a column similar to the high-velocity component (see also \citealp{Zheng20}). We apply a factor $2$ correction to the S03 median column, and estimate the MW nominal value and $1-\sigma$ error as $\novi = 1.9~(0.85-4.1) \times 10^{14}$~\cmc.

  \bgroup
 \def\arraystretch{1.5}
 \begin{table}
 \centering
 	\caption{Milky Way CGM observations}
 	\label{tab:obs_prop}
 		\begin{tabular}{| l || c | c |}
 			\midrule
  			Observable 				        &   Value ($1-\sigma$)              & Sources \\
  			\midrule
  			$P_{\rm th}/\kb$ 				& 	 $1000-3000$~K~\cmv             & (a) (b) \\
  			                 				& 	 $500-1300$~K~\cmv              & (c)     \\
 			\dml 				            & 	 $<23$~pc~\cmv                  & (d)     \\
 			                                & 	 $23 \pm 10$~pc~\cmv            & (e)     \\
 			\dmt 				            &  	 min: $63^{+27}_{-21} \pm 9$~pc~\cmv    & (f)   \\
 			\midrule
 			\novi 			& 	 $1.9~(0.8-4.1) \times 10^{14}$~\cmc    & (g) (h) \\
 			\novii 			& 	 $1.4~(1.0-2.0) \times 10^{16}$~\cmc    & (i) (j) \\
 			\noviii 		& 	 $3.6~(2.2-5.7) \times 10^{15}$~\cmc    & (k)     \\
 			\novii/\noviii 	& 	 $4.0~(2.8-5.6)$                        & (l)     \\
 			
 			\midrule
            \multicolumn{3}{|l|}{References:(a) \cite{Wolfire03} (b) \cite{Dedes10}} \\
            \multicolumn{3}{|l|}{(c) \cite{Putman12} (d) \cite{Anderson10}} \\
            \multicolumn{3}{|l|}{(e) \cite{PZ19} (f) \cite{Platts20}}  \\
            \multicolumn{3}{|l|}{(g) \cite{Savage03} (h) \cite{Zheng15}}  \\
            \multicolumn{3}{|l|}{(i) \cite{Bregman07} (j) \cite{Fang15}}  \\
            \multicolumn{3}{|l|}{(k) \cite{Gupta12} (l) \cite{FSM17}}  \\
 			\bottomrule
 	\end{tabular}
 \end{table}

\section{Results}
\label{sec_res}

In \S~\ref{sec_fsm} we described the parameter space we explore, defined by the non-thermal presssure support parameters, \alpcgm~and \vturb~(see Figure~\ref{fig:fsm_space}). We highlighted three specific parameter combinations, or models\footnote{~For the rest of this section, we refer to these combinations as ``models", although as we show next, we vary their gas masses and metallicities. This is different from the terminology in \citetalias{FSM20}, where the presented fiducial model had a single gas mass and mean metallicity.}, in this space, with negligible, moderate, and significant amounts of non-thermal pressure. They are labeled as thermal, standard and non-thermal, respectively, and their parameters are summarized in Table~\ref{tab:mod_prop}. We now explore these alternate \citetalias{FSM20} models, using inputs from our SAM galaxy sample.

The fiducial parameter set presented in \citetalias{FSM20} reproduced the MW CGM \ovii~and \oviii~measurements. Our main goal in this section is not to find the parameters best fitting the observations, but to connect FSM20 to the SAM. As we shall show, our analysis does place some constraints on the CGM mass, the amount of non-thermal pressure, and the gas profile shape in the MW. However, more importantly, it enables us to better understand how the observables behave, and provide tools for additional comparisons, to other data or models.

\subsection{Gas Distributions}
\label{subsec_res_prof}

\begin{figure*}[t]
\includegraphics[width=0.99\textwidth]{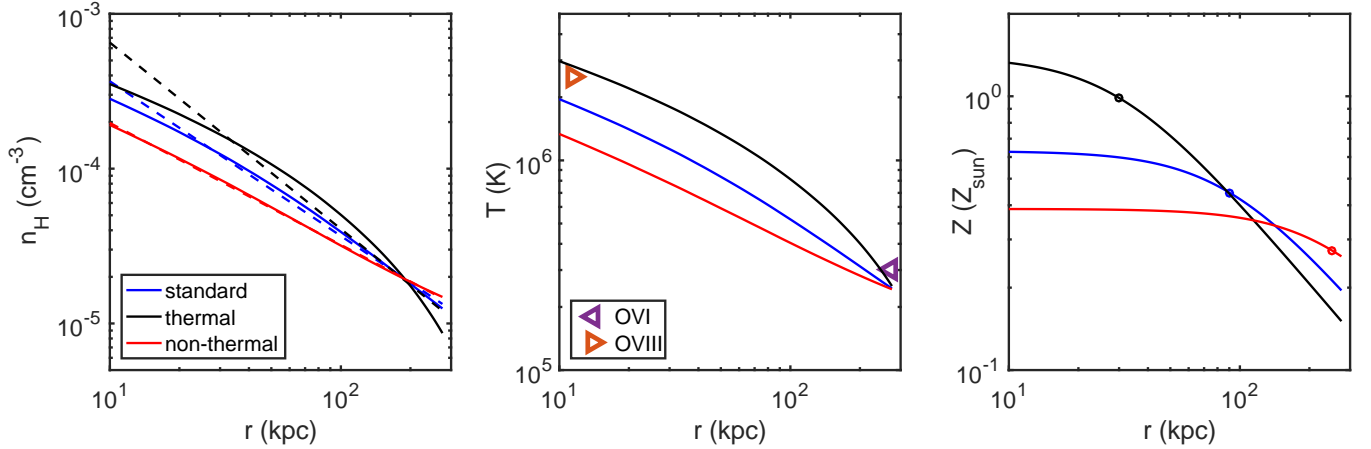}
 \caption{Profiles for the gas density (left), temperature (middle) and metallicity (right), for models with negligible, standard, and dominant non-thermal pressure support (solid black, blue and red curves, respectively), with $\mgas \sim 6 \times 10^{10}$~\msun, and $\zmean=0.3$ (see \S~\ref{subsec_res_prof}). {\bf Left:} The gas density can be approximated by a power-law, shown by the dashed curve for each profile, with slopes between $0.8$ and $1.2$. {\bf Middle:} The gas (thermal) temperature scales as $T \propto n^{2/3}$. The left- and right-pointing arrows mark the temperatures at which \ovi~and \oviii, respectively, peak in CIE. {\bf Right:} The markers show the metallicity length scales, ranging from $30$ to $250$ kpc.  (See Table~\ref{tab:mod_prop} for full model parameters.)}
  \label{fig:res_prof}
\end{figure*}

Figure~\ref{fig:res_prof} shows the gas density, temperature and metallicity profile shapes as functions of radius for our three models, for a galaxy with $\mgas \sim 6 \times 10^{10}$~\msun, and $\zmean \sim 0.3$, selected for illustration purposes. For other objects in the sample, the gas density and metallicity normalizations vary to allow for different CGM masses and mean metallicities. The normalization of the temperature profiles does not change. In the Appendix we show additional gas properties as functions of radius, such as the cooling rates, ion fractions, and ion densities.

For the thermal model, we set $\alpcgm=1.1$ and $\vturb=0$. The resulting gas density profile is shown by the solid black curve in the left panel of Figure~\ref{fig:res_prof}. This profile can be approximated by a power-law function with a slope of $a_n = 1.2$. The fit is shown by the black dashed curve, and its accuracy is between  $-20\%$ and $+50\%$ at radii beyond $15$~kpc, and a factor of $\approx 2$ at radii close to \rsun. The gas temperature profile is plotted by the black solid curve in the middle panel of Figure~\ref{fig:res_prof}. As described in \S~\ref{subsec_fsm_prop}, as a result of the polytropic EoS adopted in \citetalias{FSM20}, the shape of the temperature profile is related to the density profile, with $T_{\rm th} \propto r^{-2a_n/3}$, and the (thermal) temperature at the inner boundary is $T_{\rm th}(\rsun) \approx 3.1 \times 10^6$~K.

The parameters of the standard model are identical to those of the fiducial parameter set presented in \citetalias{FSM20}, with $\alp = 2.1$ and $\vturb=60$~\kms, resulting in $a_n = 1.0$. The profiles for this model are shown by the blue curves. For the third model we allow for significant non-thermal support, with $\alpcgm \approx 3$ and $\vturb = 100~\kms$. This leads to flatter density and temperature profiles, shown by the red curves, with $a_n = 0.80$ (accurate to within $5\%$), and $T_{\rm th}(\rsun) \approx 1.4 \times 10^6$~K. 

The right panel of Figure~\ref{fig:res_prof} shows the metallicity profiles we adopt. We choose distributions with metallicity scale lengths of $r_Z = 30$, $90$, and $250$~kpc for the thermal, standard, and non-thermal models, respectively, and these are shown by the markers on the curves\footnote{~In coupling steep (flat) gas density profiles with steep (flat) metallicity distributions we make an implicit simplifying assumption that metals follow gas. One can imagine other physical scenarios, which may be explored in future works.}.

Finally, the arrows in the middle panel show the temperatures at which the \ovi~and \oviii~ions peak at CIE, in purple and orange, respectively. This, together with the density and metallicity distributions, allows us to estimate how the column densities of each ion vary between models. For example, since the temperature at the outer boundary is fixed and close to the value at which \ovi~peaks, this ion is abundant in all three models shown here. However, in the non-thermal model, the gas temperature is closer to the \ovi~peak for a wider range of radii than in the standard and thermal models. Moreover, the gas densities and metallicities are higher there, leading to higher total \ovi~columns. For the \oviii, on the other hand, the gas temperature in the central region of the non-thermal model is below the value at which this ion peaks, resulting in lower column densities. We address the behaviour of the oxygen ion columns in detail in \S~\ref{subsec_res_oxy}, and now we start by describing the basic gas properties.

\subsection{Gas Pressure, Dispersion Measure, and Thermal Energy}
\label{subsec_res_prop}

\begin{figure*}[t]
\includegraphics[width=0.99\textwidth]{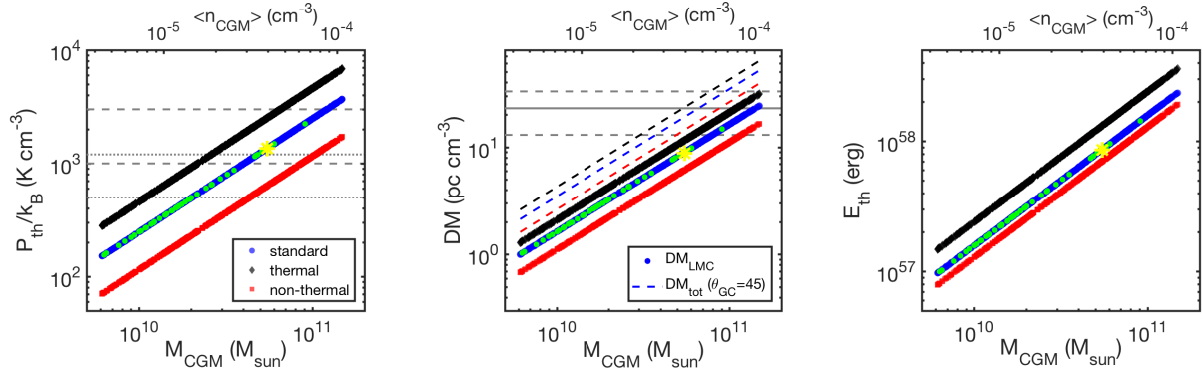}
 \caption{Gas pressure (left), dispersion measure (middle), and thermal energy (right) as functions of \mgas~or mean CGM density (top axis). The thermal, standard, and non-thermal models are shown by the black diamonds, blue circles, and red squares, respectively. The green circles show the MW-like sample, and the yellow marker is the FSM20 fiducial model~(see~\S~\ref{subsec_sam_samp}). {\bf Left:} The thermal pressure at the solar circle ($\rsun=8.5$~kpc). The horizontal grey lines (dashed and dotted) show estimates from HI clouds above the Galactic disk (see \S~\ref{sec_mw_obs}). {\bf Middle:} The DM to the LMC (markers) can be compared to pulsar observations (grey horizontal lines). The dashed lines show the total DM for an angle of $\tgc = 45^{\circ}$, and can be measured with FRBs. {\bf Right:} The total thermal energy (see details in \S~\ref{subsec_res_prop}).}
  \label{fig:res_prop}
\end{figure*}

Figure \ref{fig:res_prop} presents the gas pressure at the inner boundary, $P_{\rm th}(\rsun=8.5$~kpc), the dispersion measure, DM, and the total thermal energy of the corona, $E_{\rm th}$. The thermal, standard, and non-thermal models are plotted by black diamonds, blue circles and red squares, respectively, here and in the next figures in this section. The upper axis shows the mean gas density corresponding to the CGM mass in the bottom axis, with $\nmean = 7.2 \times 10^{-5}$~\cmv~for~$\mgas = 10^{11}$~\msun.

For a given density profile, the pressure, DM and $E_{\rm th}$ are independent of the gas metallicity, and scale linearly with the CGM total mass and mean gas density. However, each property behaves differently with the gas density profile shape. For example, the gas thermal pressure at \rsun, shown in the left panel, depends on the gas density and temperature. Both quantities are higher (lower) in the thermal (non-thermal) profiles, leading to a factor of $\sim 4$ difference in pressure between the two models. The DM (middle panel) depends on the gas density only, resulting in a smaller, factor of $\sim 2$, variation. The total thermal energy (right panel) is a function of the gas temperature and density, but integration over the entire profile erases some of the differences and gives a factor of $\sim 2$ variation. $E_{\rm th}$ is used to calculate the mean gas cooling times and mass accretion rates, which we address in \S~\ref{subsec_res_cool}.

The gas pressure in the models can be compared to the estimates from observations of High Velocity Clouds (HVCs) above the Galactic disk (see Table~\ref{tab:obs_prop}). \cite{Wolfire03} estimated a range of $1000-3000$~K~\cmv, marked in the plot by the horizontal grey dashed lines. For the standard model, objects with $\mgas > 4 \times 10^{10}$~\msun~are within this range. The pressures in the non-thermal model are lower and only the highest gas masses, $\mgas \sim 10^{11}$~\msun, have $P/\kb > 1000$~K~\cmv. The thermal model is consistent with the observationally-estimated range for $2 \times 10^{10}~\msun < \mgas < 10^{11}$~\msun. \citet{Putman12} estimated lower pressures, $\sim 500-1300$~K~\cmv~(shown by the horizontal dotted lines), shifting the mass constraints down by a factor of $\sim 2$.

In the middle panel, the markers show the dispersion measure to the LMC, at $d \approx 50$~kpc. The horizontal solid and dashed lines show the value inferred by \cite{PZ19}~(see Table~\ref{tab:obs_prop}). For our standard model, $\mgas > 8 \times 10^{10}$~\msun, and for the thermal model, $\mgas > 5 \times 10^{10}$~\msun. Both are consistent with the PZ19 estimate within $1-\sigma$. The non-thermal model has $\dml < 13$~\cmv~pc. However, if ${\rm DM} = 23$~pc~\cmv~is an upper limit to the LMC, as argued by \cite{Anderson10}, our models are consistent with it for all gas masses and profile shapes.

The total DM in our models, out to $\rcgm = 283$~kpc, depends on the angle from the Galactic Center (GC). However, the variation with angle is small, and \dmt~decreases by $\sim 25\%$ at $180^{\circ}$ (see Figure~12 in \citetalias{FSM20}). The dashed lines in the middle panel of Figure~\ref{fig:res_prop} show the DM for sightlines at $\tgc=45^{\circ}$, as a measure of the maximal DM outside the Galactic disk, and they are a factor of $\approx 2.1$ higher than \dml. These values can be compared to constraints from FRB DMs, after subtraction of the host and cosmic web contribution. \cite{Platts20} estimate ${\rm min[\dmt]} = 63^{+27}_{-21} \pm 9$~pc~\cmv, and we achieve similar values only with the highest CGM masses, with $\dmt \geq 40$~pc~\cmv~for $\mgas \geq 10^{11}$~\msun~in the thermal and standard models. We discuss this tension in \S~\ref{subsec_dis_obs}.

\subsection{Oxygen Column Densities}
\label{subsec_res_oxy}

We now examine how the gas mass, the mass-metallicity relation, and the gas distribution affect the columns of high oxygen ions measured by an observer at the solar circle looking outwards. We do this in two steps. First, in \S~\ref{subsec_res_oxy_p1}, we focus on the behaviour of the oxygen columns as a function of the gas mass in the standard model. We also compare the results using the mass-metallicity relation from the SAM to those using instead a simple analytic relation, to examine the effect of photoionization on the oxygen ions. Second, in \S~\ref{subsec_res_oxy_p2}, we examine the variation with gas distribution. We then compare the models to the oxygen columns measured in the MW in \S~\ref{subsec_res_oxy_p3}.

\subsubsection{Dependence on CGM Mass}
\label{subsec_res_oxy_p1}

We construct models with the gas mass and mean metallicity from the SAM and the standard FSM20 model parameters, and extract the oxygen column densities. In \citetalias{FSM20} the columns are functions of the angle from the GC, $\tgc$, and we take the means for angles between $40$ and $140$ degrees, similar to the lines of sight probed by QSOs in the MW. The columns are plotted in Figure~\ref{fig:res_oxy_p1} as functions of CGM mass, for \ovi, \ovii~and \oviii~(left, middle and right panels, respectively). The blue markers show the individual objects, the thick blue solid curves show power-law fits, to guide the eye, and the green markers show the MW-like sample. The horizontal grey lines show the values measured in the MW and their $1-\sig$~ranges, and we compare these to our models in \S~\ref{subsec_res_oxy_p3}.

To better understand the effects of the gas density and metallicity on the oxygen columns, we consider a simple case, in which the total mass of metals in the CGM is constant. This leads to $\zm \propto M_{\rm CGM}^{-1}$, steeper than the $M-\zm$ in the SAM, and results in a simple behaviour of the columns with gas mass. This allows us to isolate the effect of photoionization on the oxygen columns, and test whether the measured oxygen columns can be reproduced with low-mass, highly-metal-enriched CGM. Models with $\zm \propto M_{\rm CGM}^{-1}$~are shown in Figure~\ref{fig:res_oxy_p1} by the cyan solid thin curves. The CGM mean metallicity in these models is determined uniquely, and has no scatter at a given gas mass. Since we are interested in the behaviour of these models with CGM mass, the metallicity normalization is not important. We set $M_{\rm metals} = 2.0 \times 10^{8}$~\msun, corresponding to $\zmean = 0.2$ at $\mgas = 10^{11}$~\msun, close to the value of the power-law fit to the SAM $M-\zm$ (see Figure~\ref{fig:galcgm2}).

\begin{figure*}[t]
 \includegraphics[width=0.99\textwidth]{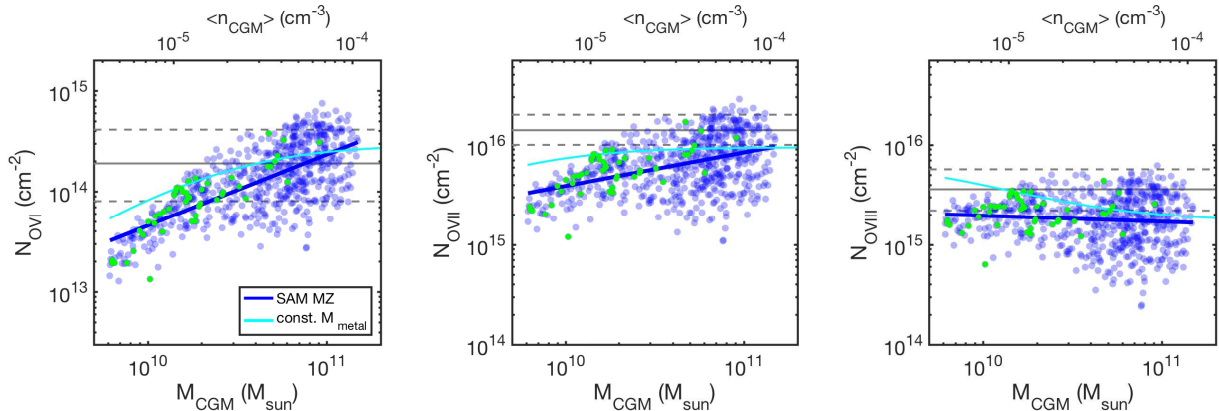}
\caption{The \ovi~(left), \ovii~(middle) and \oviii~(right) oxygen column densities for the standard model. Blue markers show models with the SAM metallicities, and the blue curve is a power-law fit to these results. The cyan curves show models assuming a constant metal mass (see \S~\ref{subsec_res_oxy_p1} for details), and the deviations from horizontal lines are due to photoionization, especially significant at low mean gas densities. The \ovi~(\oviii) is removed (created) by PI and the total column densities are reduced (increased) at low gas masses. The effect of PI on the \ovii~is small, and the cyan curve in the middle panel is close to flat. The grey lines show the values measured in the MW and their $1-\sigma$ error ranges. For the SAM metallicities, the observed \novi~and \novii~are consistent with CGM masses above $\sim 2 \times 10^{10}$~\msun. \noviii~is close to constant and does not provide a constraint on \mgas.}
  \label{fig:res_oxy_p1}
\end{figure*}

In our standard model, the gas temperature profile is fixed for any \mgas. In collisional ionization equilibrium (CIE), the mean ion fractions are independent of the gas mean density and the oxygen column densities scale as the product of the total hydrogen column (or gas mass) and mean metallicity. For models with a constant metal mass, this results in a constant column density, independent of the gas mass. The FSM20 model includes photoionization (PI) by the MGRF. This introduces a dependence of the ion fraction on the gas density. In Figure~\ref{fig:res_oxy_p1}, the effect can be estimated easily by comparing the deviations of the thin cyan curve, which includes PI, from constant column densities. In the Appendix we present individual radial ion fraction profiles for models with low and high CGM masses, and discuss them in more detail. 

At large radii, the gas temperature is close to the value at which the \ovi~peaks in CIE, $T \approx 3 \times 10^5$~K (see middle panel of Figure~\ref{fig:res_prof}). PI is then a removal mechanism for the OVI, and the effect is stronger at lower (mean) CGM densities. As result, the \ovi~column density, plotted in the left panel of Figure~\ref{fig:res_oxy_p1}, has a steeper slope as a function of CGM mass than the density-metallicity product, as the cyan thin curve shows. The deviation of the total column from CIE becomes significant ($> 30\%$) at $\mgas \sim 3.5 \times 10^{10}$~\msun. At $\mgas \sim 10^{10}~\msun$, the \ovi~column is $\sim 1/3$ of its CIE value.

For \oviii, the gas temperature is high enough for collisional ionization (CI) only in the central region. At large radii, \ovii~is photoionized to \oviii. This leads to an increase in the total \oviii~column density (right panel) at low CGM masses, with $\sim 30\%$ deviation at $\mgas \sim 3.5 \times 10^{10}$~\msun, and a factor of two at $\mgas \sim 10^{10}~\msun$. The \ovii~fraction in CIE is close to unity, and the relative change due to PI is small. The total columns (middle panel) are close to their CIE behaviour, with a $20\%$ difference at $\mgas = 10^{10}~\msun$.

A useful quantity for understanding the distributions of these ions and their integrated columns is the column density length scale, $\lscale$, defined in \citetalias{FSM20} as the radius within which half of the total column forms. The length scales are independent of the mean metallicity at a given CGM mass and they are plotted by the blue markers in the bottom panels of Figure~\ref{fig:res_oxy_p2}.

The \ovi~length scales (left panel) are large, $\sim 130$~kpc, indicating that the column forms at large radii, where the gas temperature is optimal for the ion. At low mean gas densities, PI lowers the ion fraction at larger distances, a larger part of the total column forms closer to the center, and the length scale is smaller, $\approx 100$~kpc. At high \nmean, more \ovi~survives at larger radii and \lscale~increases to $\approx 150$~kpc. The \ovii~(middle panel) requires higher temperatures, and is more centrally concentrated, with scale lengths of $\sim 25-30$~kpc. Its length scale decreases at low CGM masses due to PI, similar to \ovi. \oviii~shows a more significant variation. At high mean gas densities, PI has a small effect, \oviii~only forms in the center of the halo where the gas temperature is high enough, and the length scale is small, $< 10$~kpc. At low \nmean, \oviii~is created by PI at large radii, and a larger fraction of the total column forms there, increasing \lscale~to~$\sim 20-30$~kpc.

In the FSM20 fiducial model, the total gas mass is $\sim 5.5 \times 10^{10}~\msun$. As shown above, this mass is above the maximum threshold for a significant effect from the MGRF on the total oxygen columns. Although even at this \mgas, PI does increase the \oviii~fraction at large radii, most of the column forms at small radii ($\lscale \sim 10$~kpc), and the effect on the total column is small. In this work, we consider a wide range of gas masses, and as we have shown, at low mean gas densities photoionization does have a significant effect on the total \oviii~column densities.

\subsubsection{Dependence on Profile Shape}
\label{subsec_res_oxy_p2}

Figure~\ref{fig:res_oxy_p2} shows the \ovi-\oviii~oxygen column densities (top), and their length scales (bottom panels) as functions of CGM mass, for the thermal, standard and non-thermal models with the SAM $M-\zm$ relation. The color and marker scheme is identical to Figure~\ref{fig:res_prop}, the thick curves are power-law fits, and we omit the standard model markers for clarity. 

\begin{figure*} 
 \includegraphics[width=0.99\textwidth]{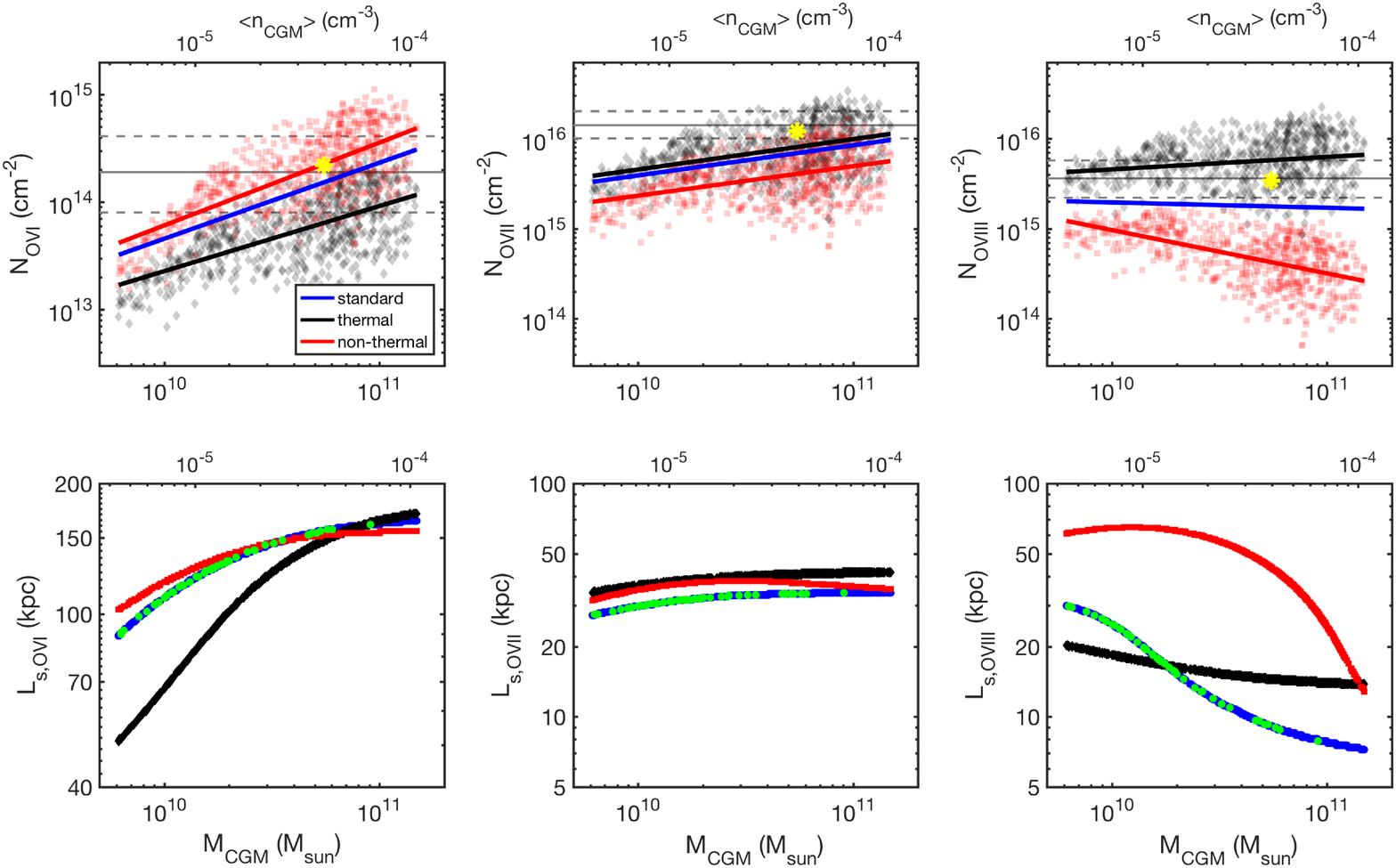}
 \caption{The oxygen column densities (top) and length scales (bottom) for the thermal, standard and non-thermal models (see \S~\ref{subsec_res_oxy_p2} for details). The color scheme is identical to Figure~\ref{fig:res_prop}. The column density length scale, \lscale, indicates the spatial distribution of each ion in the halo. \novi~(left) forms at large radii (large \lscale) and is higher in the non-thermal models. \novii~(middle) and \noviii~(right) form in the inner regions (small \lscale) and are larger in the thermal models.}
  \label{fig:res_oxy_p2}
\end{figure*}

The \ovi~columns and length scales are plotted in the left panels. At a given mass, there is less gas and metals in the thermal model (black diamonds) at larger distances from the galaxy, compared to the non-thermal model (red squares), leading to \ovi~columns that are lower by a factor of $2-4$ (top panel). The \ovi~column density length scales (bottom) are similar for the thermal and non-thermal model at high CGM masses, with $\lscale \sim 150$~kpc. This is a result of (i) the fixed gas temperature at the outer boundary, and (ii) PI having a small effect at high CGM masses. For lower CGM masses, the scale length decreases as PI removes the \ovi~at large radii. In the thermal model, with a steeper density profile, $\lscale \sim 70$~kpc at $\mgas \sim 10^{10}$~\msun, whereas the distribution is slightly more extended for the non-thermal model, with $\lscale \sim 100$~kpc. 

\ovii~and \oviii~(middle and right panels, respectively) form mostly at small radii, and the column densities of these ions are higher in the thermal models, for which the gas densities and metallicities are higher in the central part of the halo. \ovii~is abundant over a wide temperature range, and the variation with profile shape is small, a factor of $\sim 2$. The columns are weakly dependent on \mgas. The scale length (bottom middle) also varies very weakly with profile shape and CGM mass, and is between $30$ and $\sim 45$~kpc.

The \oviii~column density varies most strongly with gas profile shape, for two reasons. First, it forms mainly in the central region of the halo, where the differences between the gas densities and metallicities in the different profiles are the largest (see Figure~\ref{fig:res_prof}). Furthermore, \oviii~is sensitive to the gas temperature, and in the non-thermal model, the temperatures in the central region are $T \lesssim 1.4 \times 10^{6}$~K, below the \oviii~CIE peak temperature, at $\sim 2.5 \times 10^{6}$~K. At low CGM masses, this is compensated to some extent by formation of \oviii~through PI at larger radii. At high gas masses (and mean densities), this effect is small and the total \oviii~column density is low. This is also seen in the \oviii~length scale (bottom right), with \lscale~decreasing from $\sim 50$~kpc for $\mgas \sim 10^{10}$~\msun, to $\sim 20$~kpc at $\mgas \sim 10^{11}$~\msun. In the thermal model, the central temperatures are higher, with $T \sim 3.1 \times 10^{6}$~K. As a result, the \oviii~CIE core is more extended, and a larger fraction of the total column is formed there, leading to an increase in the column density with the mean gas density. This also results in smaller length scales, $\sim 20$~kpc, which vary only weakly with gas mass. These opposite trends with gas mass lead to an increase in the ratio of \noviii~in the thermal to the non-thermal models with mass, from $N_{\rm th}/N_{\rm nth} \sim 5$ to $\sim 20$ between $\mgas = 10^{10}$~\msun~and $10^{11}$~\msun.

\subsubsection{Comparison to Observations}
\label{subsec_res_oxy_p3}

We now compare our computed oxygen column densities to MW measurements. We focus mainly on the models with the SAM $M-\zm$ relation (Figure~\ref{fig:res_oxy_p2}) but also comment on the analytical relations presented in \S~\ref{subsec_res_oxy_p1} (Figure~\ref{fig:res_oxy_p1}). The measured columns are given in Table~\ref{tab:obs_prop}, and shown in the figures by the horizontal grey lines.

For the standard model, the \oviii~column is almost constant as a function of \mgas, and with $\noviii \sim 1.9 \times 10^{15}$~\cmc~($90\%$ in the range $0.6-4.0 \times 10^{15}$), it is close to the observed value, of $3.6 \times 10^{15}~\cmc$. The \ovi~and \ovii~columns at low \mgas~are lower than observed. To be consistent within $1-\sigma$ with the observed \ovi~column requires $\mgas > 10^{10}~\msun$. The \ovii~provides a stronger constraint of $\mgas \gtrsim 3 \times 10^{10}~\msun$.

Looking at different gas distributions, the observed \ovi~favors high masses for the thermal model, $\gtrsim 5 \times 10^{10}$~\msun. \ovii~is consistent with observations already for $\mgas \gtrsim 3 \times 10^{10}$~\msun. The \oviii~column is nearly flat and agrees with the MW value within  $1-\sigma$ for most of mass range, up to $\mgas \sim 10^{11}$~\msun.

The non-thermal model is consistent with the MW \ovi~column~over a wider mass range, down to $\sim 10^{10}$~\msun. The measured \ovii~and \oviii~columns, on the other hand, disfavor the non-thermal model, which for $\mgas \sim 10^{10}$~\msun, produces $\novii \sim 2\times 10^{15}$~\cmc, a factor of almost $10$ lower than observed. The \ovii~column increases with CGM mass, and at $\mgas \sim 10^{11}$~\msun, the discrepancy decreases to a factor of $2-3$. For the \oviii, the discrepancy is largest at high CGM masses, and at $\mgas \sim 10^{11}$~\msun, $\noviii \sim  3 \times 10^{14}$~\cmc, a factor of $\sim 10$ lower than observed. 

Since the metal ion column density is set by the density-metallicity product, we can ask whether increasing the metallicity at low CGM masses can reproduce the measured columns. For example, our constant-metals-mass models (cyan curves in Figure~\ref{fig:res_oxy_p1}) show that these are high enough for the \ovi~and the \ovii, reducing the lower limit on the CGM mass by a factor of $\sim 2$. However, the metallicities required for this are a factor of $\sim 2$ higher than in the SAM.

We can address this with the TNG100 galaxies, shown by the magenta points in Figure~\ref{fig:galcgm2}. The CGM masses in TNG do not extend to as low values as in the SAM. Fitting the $M-\zm$ with a power-law function gives a slope of $\sim 0.5$, shallower than in the SAM, and extrapolating it to $\mgas \sim 10^{10}$~\msun~results in $\zmean \sim 1$~solar, similar to the SAM metallicities at these CGM masses. The metallicities in the Santa-Cruz SAM and the TNG cosmological simulation are calculated through the growth of the stellar populations and metal enrichment in galaxies, and we conclude that the steep $M-\zm$ relation needed to reproduce the MW measurements with low CGM masses is unrealistic.

\begin{figure} 
 \includegraphics[width=0.45\textwidth]{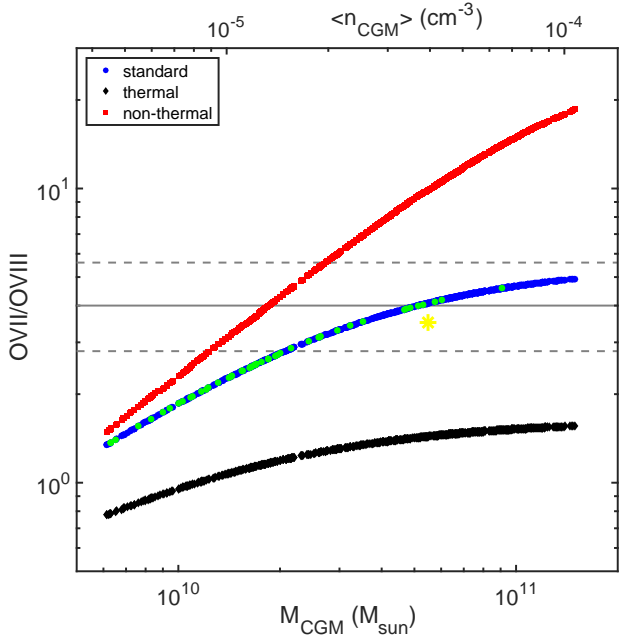}
 \caption{The \ovii/\oviii~column density ratio, with the same color scheme as in Figure~\ref{fig:res_prop} (see \S~\ref{subsec_res_oxy} for details). The ratio is independent of the gas metallicity, and sensitive to the gas distribution shape, due to the different behaviour of \noviii~with CGM mass in the thermal and non-thermal models (see top right panel of Figure~\ref{fig:res_oxy_p2}). The horizontal lines show the values inferred from MW observations, which exclude the thermal models (black), and allow non-thermal models (red) only for $\mgas \sim 1-3 \times 10^{10}$~\msun. For the standard model (blue), a wider mass range is consistent with the measured ratio.}
  \label{fig:res_oxy_rat}
\end{figure}

Finally, in Figure~\ref{fig:res_oxy_rat} we plot the \ovii~to~\oviii~column density ratio, $\chi \equiv \novii/\noviii$, which shows strong sensitivity to the gas distribution shape. For a given density and metallicity profile, the ratio is independent of the mean metallicity and does not have a scatter at a given gas mass. Overall, the \ovii~column density increases almost linearly with the gas mean density and the \oviii~does not vary significantly, resulting in $\chi$ that increases with \mgas. For the standard model, $\mgas \lesssim 2 \times 10^{10}$~\msun~are inconsistent with value estimated by \citetalias{FSM17} from observations, $\chi = 4.0~(2.8-5.6)$. The difference between the thermal and non-thermal models is dominated by the trend of the \oviii~with gas mass, and changes from $\chi_{\rm nth}/\chi_{\rm th} \sim 2-3$ at low masses to $\sim 10$ at $\mgas = 10^{11}$~\msun. The non-thermal model is consistent with the observed value of $\chi$ at CGM masses of $1-3 \times 10^{10}$~\msun~and predicts high ratios for large CGM masses. The thermal profiles produce $\chi \sim 0.1$ across the entire mass range, lower by a factor of $3-6$ than observed, and can be excluded.

Our conclusion is that given the SAM mass-metallicity relation and the standard model parameters, low CGM masses, with $\mgas<2 \times 10^{10}$~\msun, do not reproduce the oxygen column densities observed in the MW. Increasing the CGM metallicities in these objects above the values obtained in the SAM can bring the model oxygen columns into agreement with observations, but represents an unrealistic scenario in terms of galaxy evolution and metal enrichment of the CGM. Furthermore, the ratio of \ovii/\oviii~columns is independent of the gas mean metallicity and low \mgas~is inconsistent with the measured value. We discuss additional constraints from our modeling and its caveats in more detail in \S~\ref{subsec_dis_obs}.

\subsection{Cooling Rates and Times, and Mass Accretion}
\label{subsec_res_cool}

\begin{figure*} 
 \includegraphics[width=0.99\textwidth]{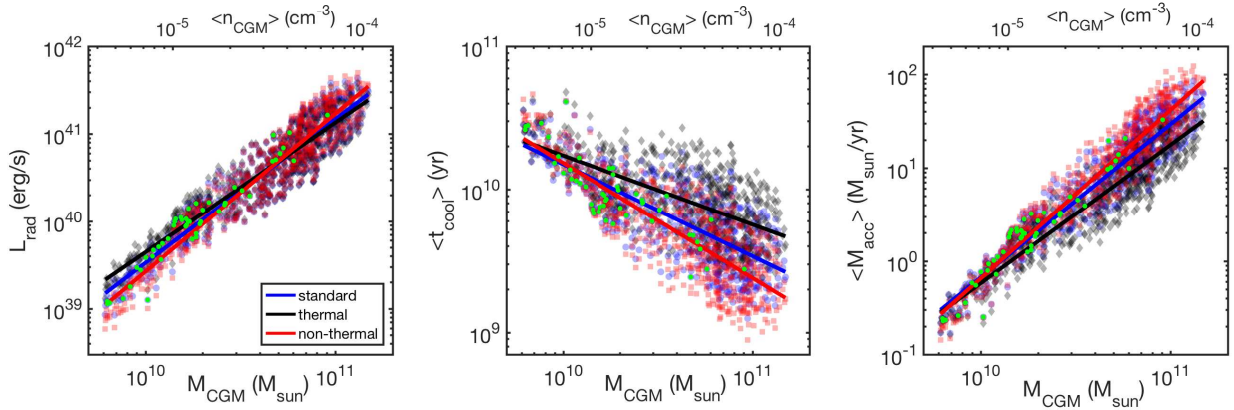}
 \caption{Gas cooling properties: the total radiative cooling rate (left), the mean cooling time (middle) and the global mass accretion rate (right). The color scheme is identical to Figure~\ref{fig:res_prop}, and the green markers show the MW subsample for the standard model. For a given model (i.e., profile shape), the cooling rate depends on the mean gas density and metallicity, with photoionization and photoheating by the MGRF reducing \Lcool~at low \mgas~(see \S~\ref{subsec_res_cool} for details).}
  \label{fig:res_cool}
\end{figure*}

Figure~\ref{fig:res_cool} shows the CGM cooling rate, cooling time and mass accretion rate. In the presence of a background radiation field, these depend on the gas density, temperature and metallicity. In the Appendix we present the radial distributions of these quantities, and here we discuss their global behaviour. The cooling rates shown in Figure~\ref{fig:res_cool} (and in the Appendix) are the gas radiative cooling rates offset by MGRF heating. They do not include energy input from galactic feedback, which we discuss in \S~\ref{subsec_dis_ebudg}.

The total cooling rate, $\Lcool$, is calculated by integrating the local cooling rates over the CGM volume, and we plot these in the left panel of Figure~\ref{fig:res_cool}, with the same color scheme as in Figure~\ref{fig:res_prop}. For each of the three model types, the temperature profile is fixed, and the total rate in the absence of the MGRF scales as $\Lcool \propto M_{\rm CGM}^2 \zmean$. At low CGM masses, photoionization and photoheating lower the cooling rates, and for the standard model (blue curve), at $\mgas \sim 10^{10}$~\msun, $\Lcool$ is reduced by a factor of $\approx 2$. The total cooling rates vary between $2 \times 10^{39}$~and $2 \times 10^{41}$~\ergs.

The cooling rates for the three models' profile shapes behave similarly, and the differences in \Lcool~at a given CGM mass are small. This is a result of several effects canceling each other. For example, in the thermal model, the inner regions have higher metallicities and higher temperatures (see Figure~\ref{fig:res_prof}), which increase and reduce the gas cooling efficiency, respectively. This leads to very similar cooling rates at high CGM masses and a factor of $\sim 2$ difference between the thermal and non-thermal \Lcool~at $\mgas < 10^{10}$~\msun. This difference is similar to the variation within each model due to the scatter in \zmean.

The mean cooling time, plotted in the middle panel, is calculated as $\tcool = E_{\rm th}/\Lcool$. The gas thermal energy scales as $\mgas$~(see right panel of Figure~\ref{fig:res_prop}), resulting in $\tcool \propto M_{\rm CGM}^{-1} \zmean^{-1}$. The suppression in \Lcool~at low gas masses leads to longer cooling times, with a factor of $\approx 2$ increase at $\mgas=10^{10}$~\msun. For the standard model, the mean cooling time is between $\sim 2 \times 10^9$ to $\sim 2 \times 10^{10}$~yr. For the thermal (non-thermal) model, \tcool~is similar at low CGM masses, and a factor of $\sim 2$ longer (shorter) at $\mgas \sim 10^{11}$~\msun. 

The global mass accretion rate is calculated by $\mcool = \mgas/\tcool$, and shown in the right panel of Figure~\ref{fig:res_cool}.  The gas mass and thermal energy cancel out, resulting in $\mcool \propto \Lcool$. For the standard model, the mass accretion rates vary between $0.5$ and $50$~\msuny.

The mean cooling times are long for all CGM masses and models, ranging from a few Gyr for high \mgas~models, to $\tcool>t_{\rm Hubble}$ at $\mgas < 2 \times 10^{10}$~\msun. This suggests that even without energy injection, the CGM is in approximate equilibrium on the timescales of galactic evolution, consistent with the assumptions of the FSM20 framework. Furthermore, as discussed in \citetalias{FSM20}, galactic feedback and dissipation of turbulent energy can offset the CGM radiative losses (see Section 4.3 there), leading to even longer cooling times. In \S~\ref{subsec_dis_ebudg}, we examine the energy injection rates from AGN in the SAM, and show that it alone is often enough to offset CGM cooling.

\subsection{The Cooling to Dynamical Time Ratio}
\label{subsec_res_tcool}

\begin{figure}[t]
\centering
\includegraphics[width=0.45\textwidth]{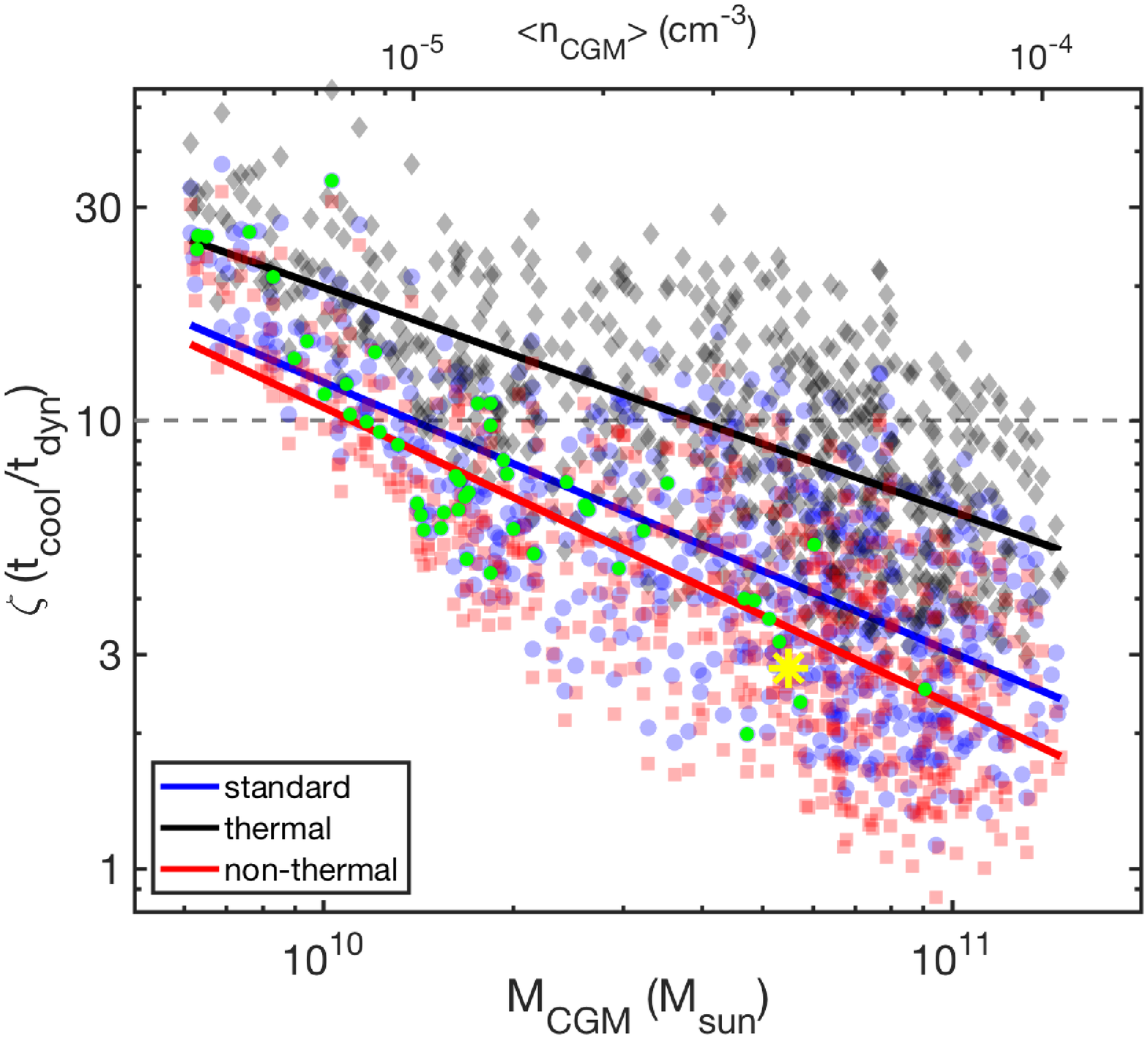} \\
\includegraphics[width=0.45\textwidth]{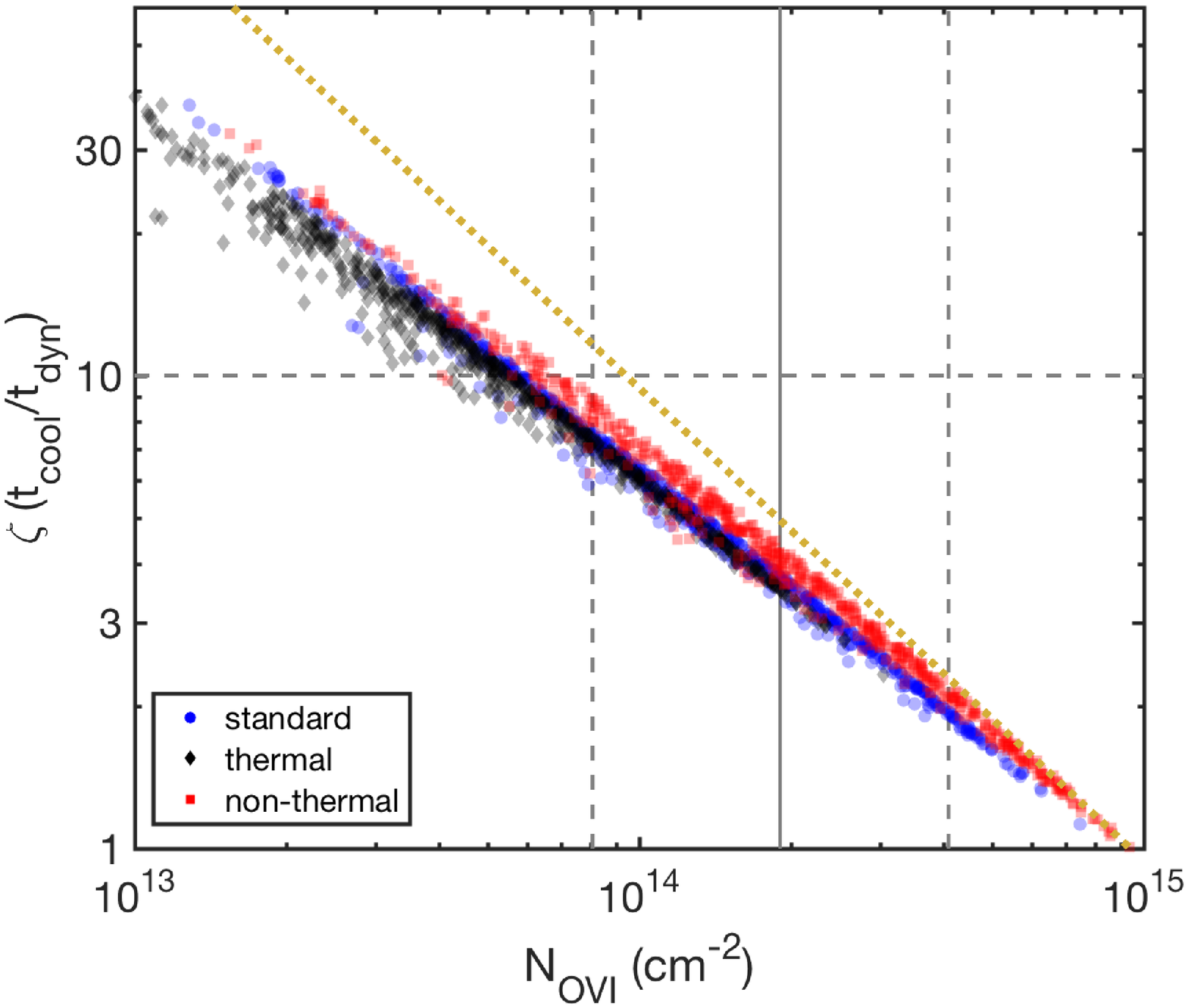}
 \caption{The mean ratio of cooling to dynamical time, $\zeta$, in the CGM (see \S~\ref{subsec_res_tcool} for details). {\bf Top:} $\zeta$~as a function of \mgas, with the horizontal dashed line showing the threshold value of $\zeta=10$. {\bf Bottom:} $\zeta$~versus the \ovi~column. The vertical grey lines show the MW estimated column, which corresponds to $\zeta \sim 4$, almost independent of the gas distribution in the CGM. The gold dotted line shows the upper limit on $\zeta$ given by Equation~\eqref{eq:tcool_lim}.}
  \label{fig:res_tcool}
\end{figure}

The gas cooling to dynamical time ratio, $\zeta = \tcool/\tdyn$, has been shown to be an interesting property of diffuse gas in halos. Idealized simulations show that gas with $\zeta<10$ is susceptible to thermal instabilities and the formation of a cool phase by precipitation and condensation \citep{Sharma12a}. Observations of galaxy clusters suggest that the intracluster medium self-regulates to have $\zeta \ge 10$~\citep{Voit15a}. The value of $\zeta$ in the CGM has been addressed by several analytic models, relating it to the \ovi~column density, gas cooling rate, and the total CGM mass \citep{FSM17,MW18,Stern18,Voit19,FSM20}. We now examine this ratio in the models presented here.

First, we consider the dependence of $\zeta$ on \mgas. As shown in the Appendix, the cooling time is almost constant with radius for high mass models and increases approximately linearly at low \mgas, due to the effect of photoheating and photoionization suppressing the gas cooling efficiency (see Figure~\ref{fig:app1}). The halo dynamical time for the gravitational potential we use \citepalias{Klypin02} can be approximated as $\tdyn \propto r^{1.22}$ (see Equation~25 in \citetalias{FSM20}). This results in $\zeta$ that is high in the inner parts of the corona, and decreases outwards. The top panel of Figure~\ref{fig:res_tcool} shows the volume-weighted mean ratio of cooling to dynamical time in the outer CGM, at $r>50$~kpc, as a function of \mgas. We use the same color scheme from Figure~\ref{fig:res_prop}, and the thick curves are power-law fits shown to guide the eye. The dashed horizontal line shows the threshold value of $\zeta = 10$. 

For the standard and non-thermal models, requiring $\zeta=10$ implies CGM masses of $\sim 1-2 \times 10^{10}$~\msun, masses that were shown in \S~\ref{subsec_res_oxy_p3} to be inconsistent with the MW oxygen column density measurements. For the thermal model, $\zeta = 10$ can be achieved with $\mgas \sim 4 \times 10^{10}$~\msun. The reason for this is that for steeper profiles, the densities and metallicities at large radii are lower, and lead to longer cooling times (see middle panel of Figure~\ref{fig:res_cool}). However, also for the thermal model, this CGM mass produces \ovi~and \ovii~column densities that are a factor of $\sim 2$ lower than MW observations (see Figure~\ref{fig:res_oxy_p2}).

To better demonstrate this point, the bottom panel of Figure~\ref{fig:res_tcool} plots $\zeta$ as a function of the \ovi~column density. This shows that in warm/hot gas, \ovi~columns above $10^{14}$~\cmc~require $\zeta<10$, almost independent of the density profile shape. The vertical lines show the MW \ovi~column (solid) and its $1-\sig$ error range (dashed). Models with $N = 1.9 \times 10^{14}$~\cmc~have $\zeta \approx 4$. We note that for the metallicity range we examine, the $\tcool-\novi$ relation is also nearly independent of the gas metallicity.

\citetalias{FSM20} derive an upper limit for $\zeta$ as a function of \novi~measured by an external observer at a given impact parameter, and apply it to the COS-Halos data. We modify this limit for an observer inside the galaxy\footnote{~This differs from the limit in \citetalias{FSM20} by a geometric factor due to the different location of the observer measuring the OVI column density.} and find that it is given by
\begin{equation}\label{eq:tcool_lim}
    \zeta \lesssim 4.7 \left( \frac{\novi}{2 \times 10^{14}~\cmc} \right)^{-1} \left( \frac{\rcgm}{280~{\rm kpc}} \right)^{-0.22} ~~~.
\end{equation}

The limit is plotted by the gold dotted line in the bottom panel of Figure~\ref{fig:res_tcool}, and the values of $\zeta$ in our models are consistent with it. For the MW column, Equation~(\ref{eq:tcool_lim}) gives $\zeta \le 5$ and to be consistent with $\zeta \sim 10$ requires $\novi <6 \times 10^{13}$~\cmc, below the $1-\sig$ range of the MW column.

\subsection{Predictions for Additional Metal Ions}
\label{subsec_res_pred}
\begin{figure*}[t]
\centering
\includegraphics[width=0.9\textwidth]{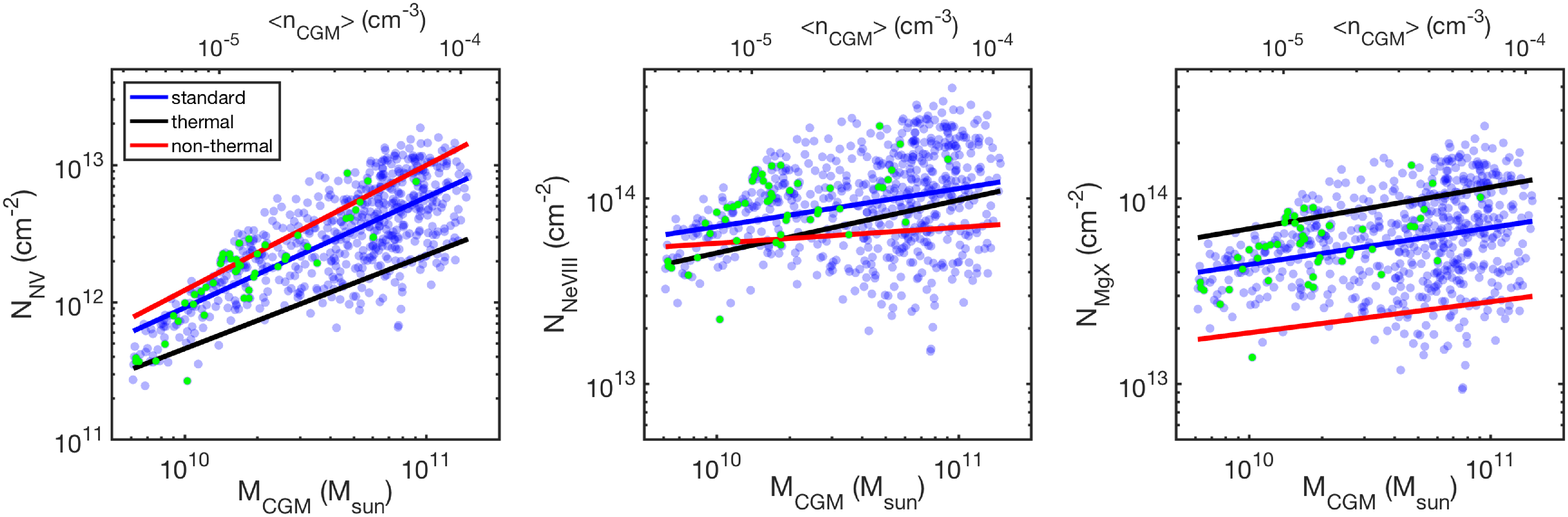}
 \caption{Column density predictions for \nv~(left), \neviii~(middle), and \mgx~(right). The three ions probe different gas temperatures, between $\sim 2.0 \times 10^5$ and $1.2 \times 10^6$~K, and behave differently with gas mass and density profile shape (see details in \S~\ref{subsec_res_pred}). \nnv~behaves similar to \novi, with the columns in the non-thermal model higher than in the thermal, but overall lower than \novi~by a factor of $>10$. \nneviii~is not very sensitive to the gas mass or distribution shape. \nmgx~forms at small radii, where the CGM temperatures are higher, and is larger in the thermal model.}
  \label{fig:res_pred}
\end{figure*}

We now provide predictions for absorption measurements of a few additional metal ions: \nv, \neviii, and \mgx. All three are Li-like ions, similar to the \ovi. However, they probe different gas temperatures, leading to different behaviour of the column densities with profile shape and gas mass, shown in Figure~\ref{fig:res_pred}.

\nv~(left panel) behaves similarly to the \ovi~(see \S~\ref{subsec_res_oxy_p2}) - the columns increase with gas mass with a power-law slope of slightly above unity, and they are higher in the non-thermal model. However, the absolute \nv~columns are low, for two reasons. First, the solar abundance of nitrogen is $6.76 \times 10^{-5}$ \citep{Asplund09}, a factor of $\approx 7$ lower than the oxygen abundance. Furthermore, in CIE, the \nv~fraction peaks at $T \sim 2 \times 10^{5}$~K, below the CGM temperature at the outer boundary of our models. This leads to \nv~columns a factor of $\sim 30$ lower than the \ovi, with $\nnv \sim 3-8 \times 10^{11}$~\cmc~at low CGM masses, and $\nnv \sim 0.1-2 \times 10^{13}$~\cmc~for high \mgas.

The middle and right panels shows the column densities of \neviii~and \mgx. While their elemental abundances are $5-10$ times lower than oxygen, these two ions reach peak fractions at $T \sim 7 \times 10^{5}$ and $\sim 1.2 \times 10^{6}$~K respectively. These temperatures are prevalent in our models, leading to column densities that are close to the \ovi, with $N \sim 10^{14}$~\cmc. The \neviii~column is not sensitive to the profile shape and varies weakly with \mgas, resulting in a relatively narrow range, of $0.4-2 \times 10^{14}$~\cmc. \mgx~forms mainly in the central region of the CGM, where the temperatures are high (similar to the \oviii), and its column density in the thermal model is a factor of $\sim 3-4$ higher than in the non-thermal. The column variation with \mgas~is weak, with a slope of $\sim 0.2$. Both ions have absorption features at $\lambda <100$~nm and may be observable in the MW with the next generation of UV space telescopes.

\section{Discussion}
\label{sec_dis}

\subsection{The MW CGM Distribution and Mass}
\label{subsec_dis_obs}

One of the goals of this work is to compare the predictions of the FSM20 model framework to observations of the warm/hot gas in the MW. \citetalias{FSM20} presented a single parameter combination, chosen to fit the MW \ovii~and \oviii~columns and other data. Here we expand on this by varying the gas distribution shape and the CGM mass, and examining models with physical mass-metallicity combinations, produced by a SAM. We now briefly summarize the results described in \S~\ref{subsec_res_prop} and \S~\ref{subsec_res_oxy_p3} and discuss possible caveats.

Comparing our models to current observational data we ask two questions. First, what gas distributions best reproduce the data? The thermal model underpredicts the \ovii/\oviii~ratio and produces low \ovi~columns. The non-thermal model has low pressures near the Galactic disk and low dispersion measure (DM). The \ovii~and \oviii~column densities are also lower than observed, by a factor of $\sim 2$ or more, and for CGM masses above $3 \times 10^{10}$~\msun, it overproduces the \ovii/\oviii~column ratio. The standard model is consistent with the measured value of each observable (DM, OVI, etc.) over a wider CGM mass range than the thermal or non-thermal models, and with a larger set of observables overall. This result supports the choice of parameters made in \citetalias{FSM20}, with similar amounts of thermal and non-thermal support.

Second, assuming a given model and density distribution, what CGM masses are preferred by the existing measurements? In general, CGM masses below $\sim 1-2 \times 10^{10}$~\msun~are inconsistent with the measured values of \novi, \novii, the \ovii/\oviii~ratio, \dml, and the pressure at the solar radius for any profile shape.
We now focus on the allowed CGM mass range for the standard model, and our figures allow the reader to perform this analysis for the two other models. The oxygen columns favor CGM masses above $\sim 2 \times 10^{10}$~\msun. Lower mass models are excluded by the \ovi~and \ovii~measurements, and the \ovii/\oviii~ratio. The DM to the LMC, if taken as a measurement, provides a stronger constraint - $\dml \sim 10-20$~pc~\cmv~requires $\mgas \sim 5-10 \times 10^{11}$~\msun. The \dmt~measurement by \citet{Platts20} suggests even higher CGM masses. However, there are two caveats to the current DM measurements. First, the \dml~may be an upper limit, due to possible contribution from the LMC ISM. Second, for \dmt, the existing sample of localized FRBs is not yet large enough to provide strong constraints. In a recent work, \citet{Keating20} argue that there is a significant uncertainty on the DM from the MW CGM, and that it can be as low as $10$~pc~\cmv. The detection of pulsars at larger distances from the MW (in dwarf satellite galaxies and M31) and a larger FRB sample will provide more information. We conclude that the MW CGM hot gas mass is in the range of $3-10 \times 10^{10}$~\msun, with a nominal value of $\sim 5 \times 10^{10}$~\msun.

The FSM20 framework addresses the warm/hot gas in the extended, spherical CGM, and there may be a contribution to the DM and the metal column densities from additional diffuse components in the MW. First, there is evidence from X-ray emission for a disk structure of hot gas around the Galaxy \citep{Nakashima18,Kaaret20}. \citet{Yamasaki20} construct a model for this disk component and find that it can have a significant contribution to the DM at low Galactic latitudes. We use the measurement to the LMC, at $l=280^{\circ}, b = -32.9^{\circ}$, and the electron column density in the disk in this direction will be small (see their Figures 6-7). DM from the disk at lower latitudes may reduce the tension with the high \dmt~inferred by \citet{Platts20}. For absorption observations, FSM17 estimated that while the hot disk can dominate the extended halo in emission, its contribution to the column densities of the high oxygen ions will be small (see Section 6.3 there). 

The FSM20 model also does not include CGM at $T<10^{5}$~K, traced by lower metal ions and lying in the vicinity of the disk \citep{Zheng19, Qu20}, or at larger distances (LMC and LMC-related complexes). This gas can be either (i) at $T \sim 10^4$~K, in a cooling/heating equilibrium with the MGRF, or (ii) in a transitional phase, between the hot and cool components. The contribution from these phases to the \ovii~and \oviii~columns is probably negligible: the temperature is too low for CI, and if the cool gas is in pressure equilibrium with the hot phase, or close to it, the gas densities in the former are also not low enough for PI to create significant amounts of \ovii~and \oviii. The \ovi~may, in principle, form in the ionized envelopes of clouds and complexes \citep{GS04}, evaporating clouds, \citep{GSM10}, or in mixing layers \citep{Ji19a,Fielding20a,Abruzzo21}. For the first two scenarios, \citet{GS04} and \citet{GSM10} showed that a single cloud is likely to contribute only a small fraction of the total \ovi~column measured in the MW ($\leq 10^{13}$~\cmc). We also do not expect a large contribution to the \ovi~column from gas cooling out of the hot phase, for two reasons. First, the gas mass is already accounted for in our models. Second, since the \ovi~peak fraction coincides with the peak of the cooling curve, the amount of cooling gas with high \ovi~should be low at any given time. For cool gas that heats up by mixing with the hot, measurements show low metallicities in the MW cloud complexes (HVCs and CHVCs), suggesting a low contribution to the measured oxygen column.

The \ovi~column density in our models is fairly sensitive to the CGM mass (see Figure~\ref{fig:res_oxy_p2}), suggesting it as a useful probe to constrain the MW \mgas. However, it also carries some uncertainty, due to the velocity cuts introduced in the measurements (see \S~\ref{sec_mw_obs}), and a possible contribution from phases or components not included in the FSM20 model. Improvements in CGM models and higher spectral resolution measurements may allow stronger constraints from \novi.

Finally, our results for the CGM mass can be compared to the mean density estimates inferred from the gas distribution around the LMC, and the absence of gas in MW dwarf satellite galaxies. In the standard model, a CGM mass of $5 \times 10^{10}$~\msun~gives $\nmean \sim 3 \times 10^{-5}$~\cmv, similar to the estimate of \citet{Blitz00} for the mean density inside $250$~kpc. The densities at $50-100$~kpc are $\sim 0.5-1.0 \times 10^{-4}$~\cmv, similar to the estimates by \citet{Grcevich09} and \citet{Salem15}. This result is non-trivial since the methods used to estimate the gas density are completely different. As mentioned above, there are uncertainties in our analysis due to additional gas structures and phases in the MW CGM. The CGM densities inferred from MW satellites are also uncertain, due to possible internal processes in the dwarf galaxies leading to gas loss (heating and outflows), the precise orbits of the satellites in the CGM, and possible stripping of gas outside the MW virial radius by the Local Group intragroup medium, for example (see \citealt{Putman21}). Thus, we find the agreement between the two results encouraging.

\subsection{The Ejected Gas Component}
\label{subsec_dis_mass}

\begin{figure*}[t]
\includegraphics[width=0.99\textwidth]{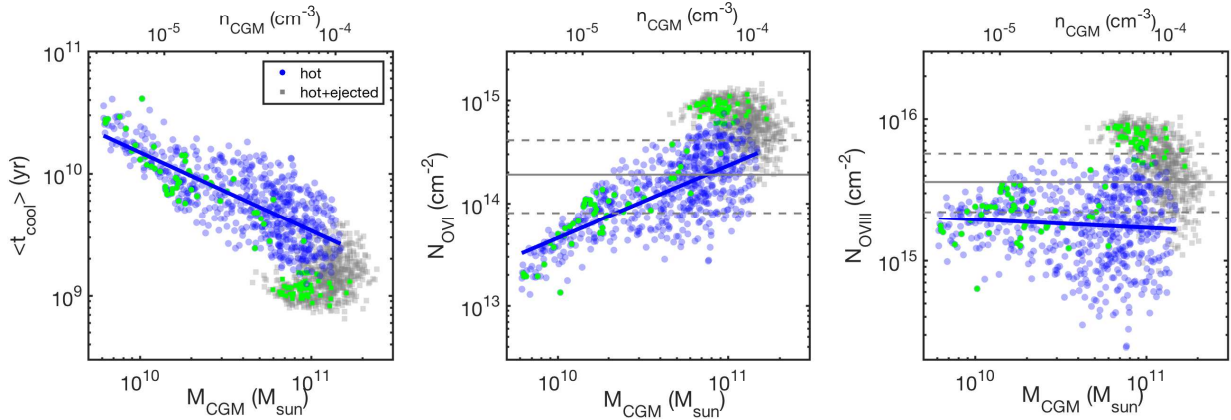}
 \caption{Massive CGM models. The grey markers show the properties of the standard model with the CGM gas and metal masses including both  the hot-halo and ejected components in the SAM (see \S~\ref{subsec_dis_mass} for details). The blue markers and curve show results when only the hot halo gas is included in the CGM, described in \S~\ref{sec_res}, for comparison. The green markers show the MW-like subsample (see \S~\ref{subsec_sam_samp}). {\bf Left}: The mean gas cooling time in the massive models is relatively short, of order $\sim \rm Gyr$. {\bf Middle and Right}: The \ovi~and \oviii~column densities in the models (markers), and the values estimated in the MW ( (horizontal grey lines)). The massive models produce \novi~that are a factor of $\sim 2-3$ higher than the measured value (middle). \noviii~(right) are also high but more consistent with observations.}
  \label{fig:dis_massive}
\end{figure*}

As described in \S~\ref{subsec_sam_frame}, the Santa Cruz SAM includes a second extended gaseous component in addition to the virialized hot CGM. This is the ejected gas reservoir, which does not become available for cooling onto the ISM for a longer timescale ($ t_{\rm re-accrete} \simeq t_{\rm Hubble}$). There are two possible physical scenarios for interpreting this reservoir --- the gas is physically ejected beyond the virial radius of the galaxy, or heated to a temperature significantly higher than the halo virial temperature, resulting in a very long cooling time. We now examine how our results change if we take the total diffuse gas mass (hot halo+ejected) as the CGM input for the FSM20 models, and then discuss the implications of our calculations for the physical interpretation of the ejected component in the SAM. We show selected results for these high mass models in Figure~\ref{fig:dis_massive}, with the hot-halo-only models for comparison.

The total gas and metal masses in the extended components are shown by the grey markers in Figure~\ref{fig:galcgm}. The full range of gas masses is relatively small, with $M_{\rm gas}$ between $5 \times 10^{10}$ and $2 \times 10^{11}$~\msun, and the mean metallicities are in the range of $0.15 < \zm < 1.5$. As shown in the right panel, the resulting $M-\zm$ relation is a very steep function of the gas mass.

For our calculation here, we adopt the parameters of the standard model (see Table~\ref{tab:mod_prop}), and the SAM gas masses and mean metallicities. The gas pressure, DM, and thermal energy scale linearly with gas mass and are identical to the properties of the standard hot-halo model (blue markers in Figure~\ref{fig:res_prop}), for the corresponding gas masses. The dispersion measure to the LMC ranges from $7.4$ to $26.0$~pc~\cmv, and the total DM is a factor of $\sim 2$ higher. The pressures at the solar radius are between $P/\kb = 1100$ and $4000$~K~\cmv, and the total thermal energy is in the range of $0.7$ to $2.5 \times 10^{58}$~ergs.

The gas cooling properties depend on the metallicity (Figure~\ref{fig:res_cool}), and the steep $M-\zm$ relation leads to high cooling rates over the range of gas masses, with $\Lcool \sim 4 \times 10^{41}$~\ergs~($\pm 0.3$~dex). This leads to short mean cooling times, $\sim 10^9$~yr, and these are plotted in the left panel of Figure~\ref{fig:dis_massive} by the grey squares. The resulting total mass accretion rates are $\sim 80$~\msuny.

The oxygen column densities are also high, with $\novi~\sim 7 \times 10^{14}$~\cmc~and $\novii \sim 3 \times 10^{16}$\cmc, outside the $1-\sigma$ range of the values measured in the MW. The \ovi~columns are shown in the middle panel of Figure~\ref{fig:dis_massive}. The \oviii~columns, plotted in the right panel, are $\sim 6 \times 10^{15}$~\cmc, closer to, but still above the nominal observed value. The mean gas densities in these models are $\left< n_H  \right> \sim 3.2 \times 10^{-5}$~\cmv. At these densities, PI has a negligible effect on the total \ovii~and \oviii~columns, and a small effect on the \ovi~column density.

To summarize, including the ejected gas mass in the CGM leads to cooling times of $\sim {\rm Gyr}$ and overproduces the \ovi~and \ovii~columns measured in the MW (see Eq. \ref{eq:tcool_lim}). One possible interpretation of this result is that in the MW, the ejected component in the SAM really does not constitute part of the CGM, and it is either ejected outside \rvir~(see also \citealp{Qu18b,Qu18a}), or it is in a hotter phase (see \citealp{Das19c,Das19a}), and thus does not contribute to the \ovi-\oviii~absorption. The scenario that some metal-enriched gas is ejected from the halo into the Cosmic Web is supported by absorption observations of the IGM, showing the presence of metals \citep{Howk09}.

Figure~\ref{fig:galcgm} shows that the ejected component constitutes a significant fraction of the total diffuse baryons in the SAM. In $\sim 65\%$ of the galaxies in our sample the ejected mass exceeds the hot halo component. In galaxies with stellar masses similar to the MW (marked by green circles), the mass of the CGM is below $10^{11}$~\msun, with only a few objects above $5 \times 10^{10}$~\msun. The high masses of the ejected gas reservoir in the SAM were recently addressed by \cite{Pandya20}, who find that the mass outflow rates in the SAM are higher than those measured in high resolution cosmological zoom-in simulations from the FIRE suite. These high mass ejection rates are required in the SAM to avoid the build-up of over-massive stellar components, which are inconsistent with the stellar mass to halo mass relation. \cite{Pandya20} suggest that modifying the SAM to include heating of the CGM by stellar winds, dubbed ``preventative feedback", will lead to lower accretion rates from the CGM. This will eliminate the need to expel large gas masses from the galaxy, and may lead to the CGM retaining higher fractions of the galactic baryons. It will be interesting to repeat our modeling once this physical recipe is added to the SAM. Another mechanism that can prevent the build-up of massive ejected reservoirs is to allow SN-driven winds to escape from the halos completely, similar to the current treatment of AGN-driven winds. However, this is more likely to happen in lower mass halos, and probably will not have a strong effect in MW-like galaxies.

\subsection{CGM Cooling/Heating Balance}
\label{subsec_dis_ebudg}

\begin{figure*}[t]
\includegraphics[width=0.99\textwidth]{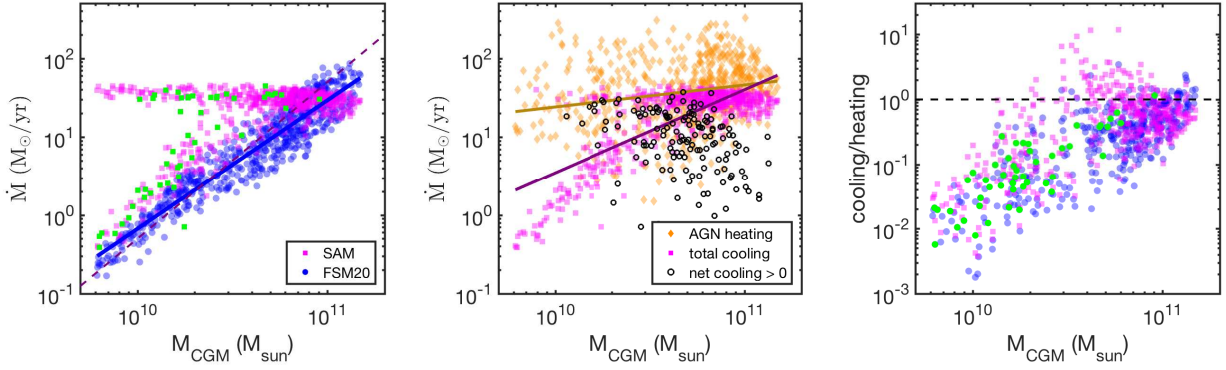}
 \caption{CGM mass cooling and heating rates in the FSM20 model and in the SAM (see \S~\ref{subsec_dis_ebudg} for details). {\bf Left:} The total cooling rate of the CGM in FSM20 (blue circles) and the SAM (magenta squares). In the SAM, the accretion mode sets the behaviour of the cooling rate with \mgas, with objects in the cold mode predicted to have a constant $\dot{M}_{\rm cool}$ within the halo dynamical time. Objects in the hot mode have $\dot{M}_{\rm cool} \propto \mgas^2$ (dashed purple curve), and their cooling rates are similar to those of FSM20. {\bf Middle:} In the SAM, cooling in the hot accretion mode is offset by AGN jet-driven heating (orange diamonds), and about $30\%$ of these objects in our sample have non-zero net cooling rates (black circles). {\bf Right:} Cooling to (AGN) heating ratio in the SAM is below unity for $\sim 70\%$ of the objects, and $\sim 90\%$ of the objects when compared to the FSM20 cooling rates.}
  \label{fig:dis_ebudget}
\end{figure*}

The FSM20 picture assumes that the CGM is in a large-scale equilibrium, and that the gas radiative losses are, at least partially, offset by heating processes. We now examine the energy budget by comparing the cooling rates in the FSM20 models to the gas cooling and heating rates in the SAM. We show the results in Figure~\ref{fig:dis_ebudget}.

The left panel shows the CGM mass cooling rates. The blue circles and solid curve are the cooling rates from the standard model presented in \S~\ref{sec_res}. The mean \mcool~is calculated as the total CGM mass divided by the mean cooling time, and for a given gas distribution, $\mcool \propto \Lcool$ (see \S~\ref{subsec_res_cool}).

The magenta markers show the cooling rates calculated in the SAM, where the CGM is assumed to follow a singular isothermal sphere (SIS) density profile with a constant metallicity. The cooling efficiencies used are from \citetalias{SD93}, assuming CIE. \citetalias{Somer08} define the cooling radius, \rcool, as the radius within which the gas cools on a timescale given by $t_c$, and $\rcool \propto t_c^{1/2}$. The total mass cooling rate from the CGM into the ISM is then given by Equation (2) in \citetalias{Somer08}:
\begin{equation}\label{eq:sam_cool}
\dot{M}_{\rm cool} = \frac{1}{2} \frac{\mgas}{t_c} \frac{\rcool}{\rvir} ~~~.
\end{equation}
\citetalias{Somer08} take the cooling timescale to be the halo dynamical time, $t_c = \tdyn = \rvir/\vvir$.

The SAM mass cooling rates show a bimodal behaviour. For one subgroup of objects in our sample, \mcool~increases with CGM mass, with a slope similar to that of the purple dashed line, scaling as $M_{\rm CGM}^2$. For these galaxies, the mass cooling rates calculated by the FSM20 model and the SAM are similar, with $\sim 1~(35)$~\msuny~at $\mgas \approx 10^{10}~(10^{11})$~\msun. For the second group of objects, \mcool~is predicted by the SAM to be approximately constant with CGM mass, with $\mcool \sim 30-40$~\msuny. Many of these objects are identified by the SAM to be in the cold accretion mode, defined by $\rcool>\rvir$, and these constitute $\sim 10\%$ of our sample. However, we find that some objects along the constant \mcool~branch have $\rcool<\rvir$, and these typically have CGM masses above $\sim 3 \times 10^{10}$~\msun. We attribute these objects to halos in transition between the cool and hot modes. In future analysis, they can be excluded by adopting a lower value for the maximal cooling radius. For example, requiring $\rcool < 0.85 \rvir$ removes most of the horizontal branch objects from our sample, and $\sim 65\%$ of the full sample survives this criterion.

In the SAM, the CGM of galaxies in the hot accretion mode can be heated by AGN radio jets. The mass heating rate is given by Equation (21) in \citetalias{Somer08}
\begin{equation}\label{eq:sam_heat1}
\dot{M}_{\rm heat} = \frac{L_{\rm heat}}{(3/2) \kb T/\mu m_p} = 
\frac{\kappa_{\rm heat} \eta_{\rm rad} \dot{m}_{\rm radio}c^2}{(3/4) \vvir^2} ~~~,
\end{equation}
where $\vvir$~is the halo virial velocity, $\dot{m}_{\rm radio}$~is the mass accretion rate in the AGN radio mode, $\eta_{\rm rad}$~is the conversion efficiency of the accreted mass to radiation, and $\kappa_{\rm heat}$~is the coupling efficiency of the radio jets with the hot gas. 

We plot these AGN jet-driven mass heating rates in the middle panel of Figure~\ref{fig:dis_ebudget} (orange diamonds) and we re-plot the SAM mass cooling rates from the left panel for objects in the hot accretion mode ($\rcool<\rvir$), for comparison (magenta squares). The thick solid lines are power-law fits plotted to show the trends in the data. The AGN heating rates have a significant scatter at a given CGM mass, with a full range of $\mheat \sim 5$ to $300$~\msuny. The fit shows a weak correlation with \mgas, with typical values of $\sim 30-50$~\msuny. For most objects, the mass cooling rates have lower or similar values. The black markers show the SAM net mass cooling rates ($\mcool-\mheat$) for systems with positive net cooling. These constitute $\sim 30\%$ of the objects in our sample. The net cooling rates have a median value of $\sim 10$~\msuny, and a scatter of $\pm 0.5$~dex. In the right panel, we plot the cooling to heating ratio for each object, both with the SAM and the FSM20 cooling rates (magenta and blue markers, respectively). The FSM20 cooling rates are lower than in the SAM, and comparing them to the AGN heating rates, only $\sim 10\%$ of the objects have positive net cooling.

This result supports the estimate in \citetalias{FSM20}, that the energy released by accretion onto the SMBH can be enough to offset the radiative cooling of the CGM (see Section~4.3 there). Furthermore, as noted earlier, the current Santa Cruz SAM does not include CGM heating from stellar feedback (but only ejection of the ISM), which can also have a significant contribution to the total heating rate \citep{Pandya21}, lowering the net mass cooling rate even further.

However, we do not argue that the radiative losses in the coronal gas must be offset exactly. How the energy injected by AGN or stellar feedback couples to the CGM and the efficiencies of these processes remain open questions, and a precise balance is not required. First, as we have shown in \S~\ref{subsec_res_cool}, for low CGM masses, the cooling time is long, comparable to the Hubble time (see middle panel of Figure~\ref{fig:res_cool}). In these cases, even if the radiative losses are not balanced, the CGM can be in approximate equilibrium. Second, for galaxies with a non-negligible SFR, the mass cooling rate of the CGM does not need to be zero. 

To demonstrate the second point for the MW, we fit the mass cooling rate calculated by FSM20 as a function of \mgas
\begin{equation}
\dot{M}_{\rm cool} = 9.4~(5.1-13.7) \times \left( \frac{\mgas}{5 \times 10^{10}~\msun} \right)^{1.64} ~ \msuny ~~~,
\end{equation}
where the range in the normalization factor represents the $1-\sigma$ scatter in the sample, $\Delta \dot{M}_{\rm cool} \pm 0.21$~dex. The total cooling rate can be written as $\dot{M}_{\rm cool} = SFR + \dot{M}_{\rm heat}$. Taking the CGM mass estimated in \S~\ref{subsec_dis_obs}, $\mgas\sim 5 \times 10^{10}$~\msun, and the MW SFR, $\sim 2$~\msuny~\citep{Chomiuk11,Lic15}, suggests a mass heating rate of $\sim 7.4$~\msuny, or $L_{\rm heat} \sim 3.8 \times 10^{40}$~\ergs. Reversing this relation and writing the CGM mass as a function of the mass cooling rate gives a lower limit on the CGM mass for a measured star formation rate
\begin{equation}
\mgas \gtrsim 2.1~(1.5-2.7) \times 10^{10} \times \left( \frac{SFR}{2~\msuny} \right)^{0.55} ~ \msun ~~~,
\end{equation}
with a range of $\pm 0.12$~dex. Equality occurs when $\dot{M}_{\rm heat} \approx 0$, if there is no energy injection into the CGM (through feedback or otherwise), or if the feedback coupling efficiency to the CGM is very low. For the MW SFR, this gives a gas mass similar to the estimates by \citet{Stern19} and \citet{Voit19}, $\mgas \sim 2 \times 10^{10}$~\msun.

To summarize this point, the FSM20 model framework provides the CGM cooling rate as a function of the gas mass and distribution. When examined together with the galaxy SFR and SMBH activity, the cooling rate can be used to infer a lower limit on the CGM mass for a given SFR or estimate the amount of feedback needed to offset the cooling for a given CGM mass. Knowing both can allow us to study how efficiently the feedback couples to the CGM or predict whether the galaxy is building or depleting its ISM reservoir.

\section{Summary}
\label{sec_summ}

In this paper we combined the Santa Cruz SAM for galaxy formation with the FSM20 model to examine how CGM observables, such as the dispersion measure and oxygen column densities, behave with the total gas mass and its spatial distribution. We test how measurements of these observables constrain these important properties of the MW CGM, and provide predictions for comparison with other models and future observations. We also explore how the radiative luminosity and cooling time behave with \mgas, and study the mass accretion rate and energy budget of the CGM.

In Section~\ref{sec_sams} we use the SAM to generate a sample of $z=0$ galaxies with MW-like halo masses, $\mvir \approx 10^{12}$~\msun. These galaxies have a wide range of stellar masses and star formation rates (Figure~\ref{fig:galprop}). The diffuse extended baryons in the SAM have two components, hot halo and ejected gas, and we consider the former to constitute the CGM in our ``standard" models. The mean metallicity of this component in the SAM galaxies is correlated with its mass, with $\zmean \propto M_{\rm CGM}^{-0.75}$, but with significant scatter at a given CGM mass (Figure~\ref{fig:galcgm}). We find that the CGM mass and metallicity of the fiducial model from \citetalias{FSM20} and those of the SAM galaxies are consistent, which is a non-trivial result.

We present the parameter space of the FSM20 model we explore in Section~\ref{sec_fsm}. Given a constant halo mass, we fix the spatial extent of the CGM and the gas temperature at the outer boundary, and vary the amount of non-thermal pressure support in the CGM, setting the shape of the gas density profile. We find that the density can be well-approximated with a power law function, with $n \propto r^{-a_n}$. The slope varies between $a_n\sim 1.25$ for distributions with only thermal pressure, and $\sim 0.75$ for models with dominant non-thermal support (Figure~\ref{fig:fsm_space}). We summarize the MW CGM observations with which we compare our models in Section~\ref{sec_mw_obs}.

We present our main results in Section~\ref{sec_res}. We use the gas masses and mean metallicities from the SAM as input parameters for the FSM20 framework, resulting in a physically-motivated exploration.  We construct models with three parameter combinations, sampling different amounts of non-thermal support, and examine how the CGM properties vary with gas mass and distribution. First, in \S~\ref{subsec_res_prof}, we present the gas density, temperature and metallicity profiles (Figure~\ref{fig:res_prof}). Then, in \S~\ref{subsec_res_prop}, we address the gas pressure and the dispersion measure. We show that the pressures estimated from HVCs above the Galactic disk and DM measurements disfavor low CGM masses and models dominated by non-thermal support. However, these observations currently do not provide strong constraints.

In \S~\ref{subsec_res_oxy}, we examine the columns densities of high oxygen ions, \ovi-\oviii, as observed from inside the galaxy. These depend on the gas metallicity, and at low gas densities (or CGM masses), the ion fractions are also significantly modified from their CIE values by the UV background. To isolate these effects, we construct models with a constant metal mass.  We find that (i) photoionization starts to significantly affect the total oxygen columns at $\mgas \sim 3 \times 10^{10}$~\msun, and (ii) to reproduce the observed columns with low CGM masses requires metallicities that are significantly higher than those predicted by the SAM $M-\zm$ relation (Figure~\ref{fig:res_oxy_p1}). For models with the SAM CGM masses and metallicities, we show that each oxygen ion behaves differently with profile shape and \mgas, demonstrating the strength of combining data from different ions when comparing to models (Figure~\ref{fig:res_oxy_p2}).

In \S~\ref{subsec_res_cool}, we present the CGM cooling rates, the implied cooling times, and the mass accretion rates in the absence of feedback (Figure~\ref{fig:res_cool}). We show that the total cooling rates are similar for the different profile shapes, with a difference of a factor of $<2$ between the profiles, similar to the scatter resulting from the $M-\zm$ relation. We then address the cooling to dynamical time ratio, $\zeta$, as a function of gas mass in \S~\ref{subsec_res_tcool}, and find that $\zeta > 10$ requires $\mgas < 2 \times 10^{10}$~\msun~(Figure~\ref{fig:res_tcool}). We also derive an upper limit on the cooling time as a function of the \novi, and show that the column measured in the MW requires $\zeta < 5$, similar to the value \citetalias{FSM20} estimate for the COS-Halos galaxies. Finally, in \S~\ref{subsec_res_pred} we present predictions for \nv, \neviii~and \mgx~column densities in the MW CGM (Figure~\ref{fig:res_pred}). These, together with other metal ions, can be incorporated into future versions of the SAM.

In Section~\ref{sec_dis} we address three topics. First, we summarize our constraints on the CGM mass in \S~\ref{subsec_dis_obs}, and find that the MW measurements favor CGM masses of $\sim 3-10 \times 10^{10}$~\msun, and profiles with similar amounts of thermal and non-thermal support. This is similar to the fiducial model presented in \citetalias{FSM20}. However, this extended analysis allows a better understanding of how the different observables behave with model parameters. Second, in \S~\ref{subsec_dis_mass}, we show that including the ejected component from the SAM in the CGM mass leads to (i) short cooling times, with $\tcool \sim $~Gyr, and (ii) \ovi~and \ovii~columns that are higher than the measured values by a factor of $\sim 2$ (Figure~\ref{fig:dis_massive}). One possible conclusion is that a significant fraction of the gas and metals were ejected beyond \rvir, and our Galaxy lacks $\sim 30\%$ of its baryons. 
Another option is this gas is within the halo but is at a different temperature or in a different ionization state than assumed in the simplest model scenario, and therefore cannot cool efficiently, 
possibly as a result of preventative feedback (see \citealp{Pandya20}). In this scenario the CGM has a higher mass and lower metallicity. Finally, in \S~\ref{subsec_dis_ebudg} we show that the AGN heating rates in the SAM are enough to offset the radiative losses in the CGM for a large fraction of the objects in our sample (Figure~\ref{fig:dis_ebudget}). However, we argue that the cooling rates calculated in the FSM20 framework do not need to be balanced exactly, and comparing them with the energy output of the galaxy can allow us to constrain the efficiency of CGM heating by feedback and predict the future evolution of the galaxy.

We hope that the combination of semi-analytic models for galaxy formation and detailed CGM models, as performed in this work, will improve our understanding of galaxy evolution and the CGM. We also hope that future observations will allow to use our models to put even stronger constrains on the CGM properties and better understand this important component of our Galaxy.

\begin{acknowledgements}

\vspace{-0.3cm}

We thank Yossi Cohen for sharing his catalogue of TNG100 galaxies and CGM properties. We thank Yuval Birnboim, Avishai Dekel, Orly Gnat, Ariyeh Maller, Nir Mandelker, Kartick Sarkar and Jonathan Stern for helpful discussions and suggestions during the course of this work. 

This research was supported by the Israeli Centers of Excellence (I-CORE) program (center no.~1829/12), the Israeli Science Foundation (ISF grant no.~857/14), DFG/DIP grant STE 1869/2-1 GE625/17-1, and the Mathematics and Physical Sciences (MPS) program of the Simons Foundation, and the Center for Computational Astrophysics (CCA) at the Flatiron Institute. Y.F.~thanks the CCA for hospitality. This work also benefited from discussions held during the program "Fundamentals of Gaseous Halos" at the KITP (UCSB), supported by the National Science Foundation under Grant No.~NSF PHY-1748958. RSS is supported by the Simons Foundation, and some of the calculations presented in this work were carried out on the Flatiron Institute Computing Cluster.

\end{acknowledgements}


\addcontentsline{toc}{section}{References}

\bibliographystyle{yahapj}



 \renewcommand{\theequation}{A-\arabic{equation}}
 \renewcommand{\thefigure}{A-\arabic{figure}}
 \setcounter{equation}{0}
 \setcounter{figure}{0}
 \setcounter{section}{0}
 \section*{APPENDIX - Radial CGM Profiles}  
 \label{sec_appprof}

In this Appendix we present the distribution of gas properties with radius in our CGM models, and examine how they are affected by variation in the model parameters. In Section \S~\ref{sec_res} we constructed three sets of models, with different parameter combinations setting the gas density and metallicity profile shapes (see Table~\ref{tab:mod_prop}). One is the fiducial model from \citetalias{FSM20}, and the other two cases are (i) models with only thermal support, resulting in steeper gas density profiles and (ii) models with significant non-thermal (turbulent and B/CR) support, leading to flatter profiles. The shape of the gas metallicity profile was also chosen to be steep or flat, through the gas metallicity length scale, $r_Z$. In each set we constructed models with the gas masses and mean metallicities from the SAM, with $6 \times 10^9 < \mgas < 1.3 \times 10^{11}$~\msun, and $0.1 < \zm < 2.0$. 

Figures~\ref{fig:app1} and \ref{fig:app2} show the radial profiles of different gas properties for these three parameter combinations, and for each case we show two specific models, with CGM masses of $\mgas =10^{10}$ and $10^{11}$~\msun, bracketing the gas mass range given by the SAMs. These models have $\zm \approx 0.1$ and $\zm \approx 1.0$ for the high and low CGM mass models, respectively. This results in a constant density-metallicity product, and for properties like the metal ion densities, allows us to focus on higher order effects, such as temperature and photoionization (PI). The dashed and solid curves show the low and high mass models, respectively. The color coding is identical to Figure~\ref{fig:res_prof} - the blue curves show models with the FSM20 fiducial parameters, and the black (red) - models dominated by thermal (non-thermal) pressure support.

\subsection*{G\MakeLowercase{as} P\MakeLowercase{roperties} \MakeLowercase{and} C\MakeLowercase{ooling}}

\begin{figure*}
 \includegraphics[width=0.98\textwidth]{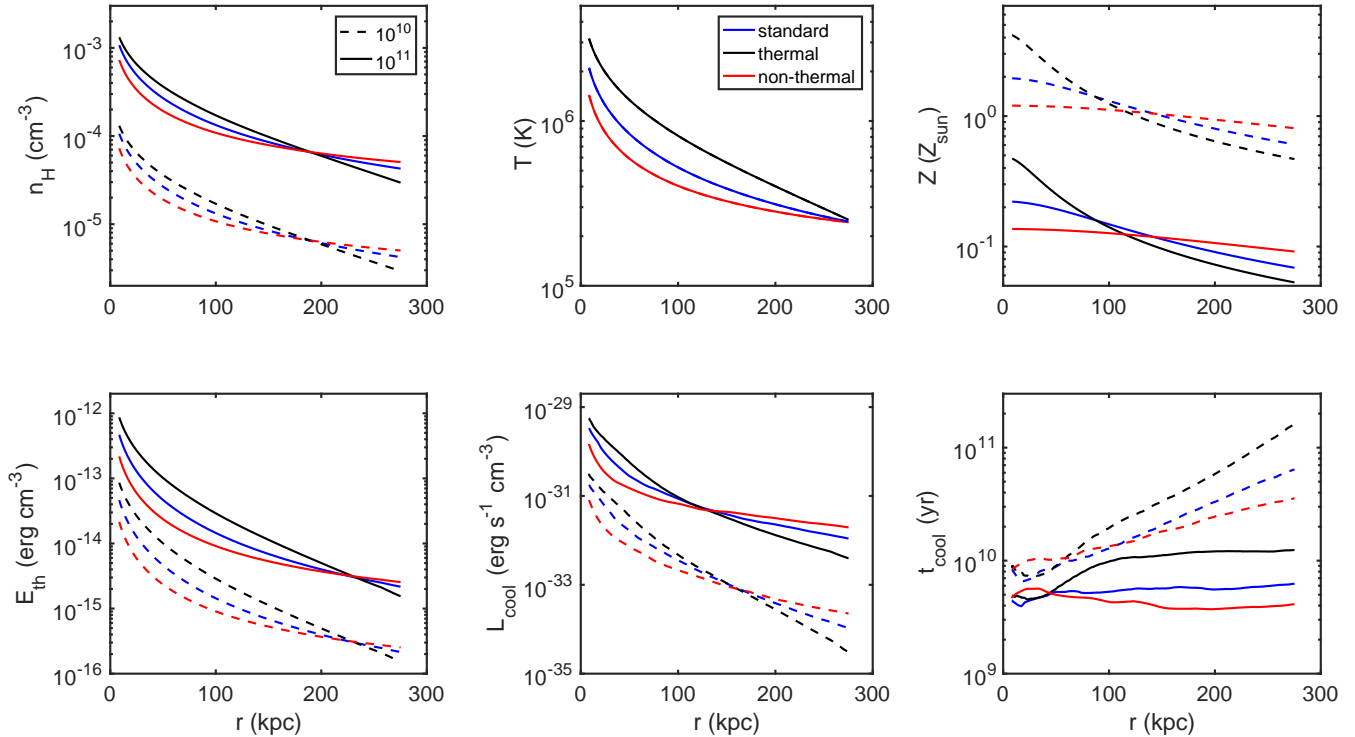}
 \caption{Gas properties versus radius. {\bf Top:} Gas density, temperature and metallicity (left, middle and right panels, respectively). {\bf Bottom:} Gas thermal energy density, radiative cooling rate and cooling time. Each panel shows the gas distributions for three models defined by the slope of their gas and metallicity profiles - fiducial (blue curves), steep (black) and flat (red). The dashed and solid curves show profiles for $\mgas = 10^{10}~\msun$ and $10^{11}~\msun$, respectively.}
  \label{fig:app1}
\end{figure*}

The top panels of Figure~\ref{fig:app1} show the gas density, temperature and metallicity profiles. The gas density profiles (left) can be approximated by power-law functions with slopes $a_n \sim 1$ (see~\S~\ref{sec_fsm} and left panel of Figure~\ref{fig:fsm_space}), and the gas mass distribution is dominated by large radii, with the profiles intersecting at $r \sim 200$~kpc. The gas thermal temperature profiles (middle) do not depend on the mean gas density and we show a single set of curves for the three profile shapes. The temperature at the outer CGM boundary is constant and set to the temperature of the MW virial shock, $2.4 \times 10^5$~K (see~\S~\ref{subsec_fsm_prop}). The gas entropy in our model is constant with radius and the temperature is related to the gas density through the adiabatic EoS, with $T \propto n^{2/3}$ for the thermal component. For steeper (flatter) gas density distributions, this results in steeper (flatter) temperature profiles and higher (lower) temperatures in the central regions of the halo. For example, the temperature at the inner CGM boundary ($r \sim 10$~kpc) is $\approx 3.1 \times 10^6$~K for the steep gas density profile, and $\approx 1.4 \times 10^6$~K for the flat density models (see also Table~\ref{tab:mod_prop}).
 
The top right panel shows the gas metallicity profiles. The models shown here have mean metallicities of $\zmean \approx 0.1$ and $1.0$~solar for the high and low mass, respectively, chosen to give a constant $M-\zm$. The metallicity profile shape is determined by the metallicity lenth scale, $r_Z$ and for our models, we couple steep (flat) gas density profiles to steep (flat) metallicity profiles, with small (large) $r_Z = 30$~($250$)~kpc. The metallicity decreases outwards, leading to a distribution of metals that is more centrally concentrated than that of gas.

The bottom panels of Figure~\ref{fig:app1} show the gas thermal energy, radiative cooling rate and cooling time profiles. The thermal energy density (left panel) is given by $E_{th} = \frac{3}{2} n \kb T$. The higher gas temperatures in the steep models result in higher local gas thermal energies at $r<250$~kpc, and higher total thermal energies (see right panel of Figure~\ref{fig:res_prop}). The local gas radiative cooling rate, given by $\Lcool \propto n^2 \Lambda(n,T,Z)$, is plotted in the bottom middle panel. In the models we examine here, regions with higher gas density also have higher metallicities and higher cooling rates. However, the gas temperatures there are also higher, and in the temperature range of our models, the cooling efficiency decreases with temperature. This leads to similar cooling rate profiles and total cooling rates for the different model sets (see left panel of Figure~\ref{fig:res_prop}). The gas local cooling time is calculated as $\tcool = E_{\rm th}/\Lcool$, and the cooling times (bottom right panel) are longer for the thermal models. We note that the high-mass models presented here have long cooling times, between $\sim 4 \times 10^{9}$ and $10^{10}$~years. This is longer than the typical times shown in Figure~\ref{fig:res_prop}, with $2 \times 10^9$~years, a result of the low mean metallicity of the high mass models shown here.

It is interesting to compare the gas cooling rates and times in the low and high CGM mass models (dashed and solid curves, respectively). At large radii the gas densties are low, photoheating by the MGRF suppresses the gas cooling efficiency and results in steeper cooling rate profiles for the low mass models of a given profile shape. For example, for the thermal profiles (black curves) the ratio of the gas cooling rate of high to low mass models is $\sim 20$ at $\sim 10$~kpc and increases to $\sim 130$ at $\sim 280$~kpc. As a result, the cooling time profiles for the high mass models are almost flat with radius (at $r>100$~kpc), and increase almost linearly with radius for the low mass models.

\subsection*{O\MakeLowercase{xygen} F\MakeLowercase{ractions} \MakeLowercase{and} D\MakeLowercase{ensities}}
\begin{figure*}
 \includegraphics[width=0.98\textwidth]{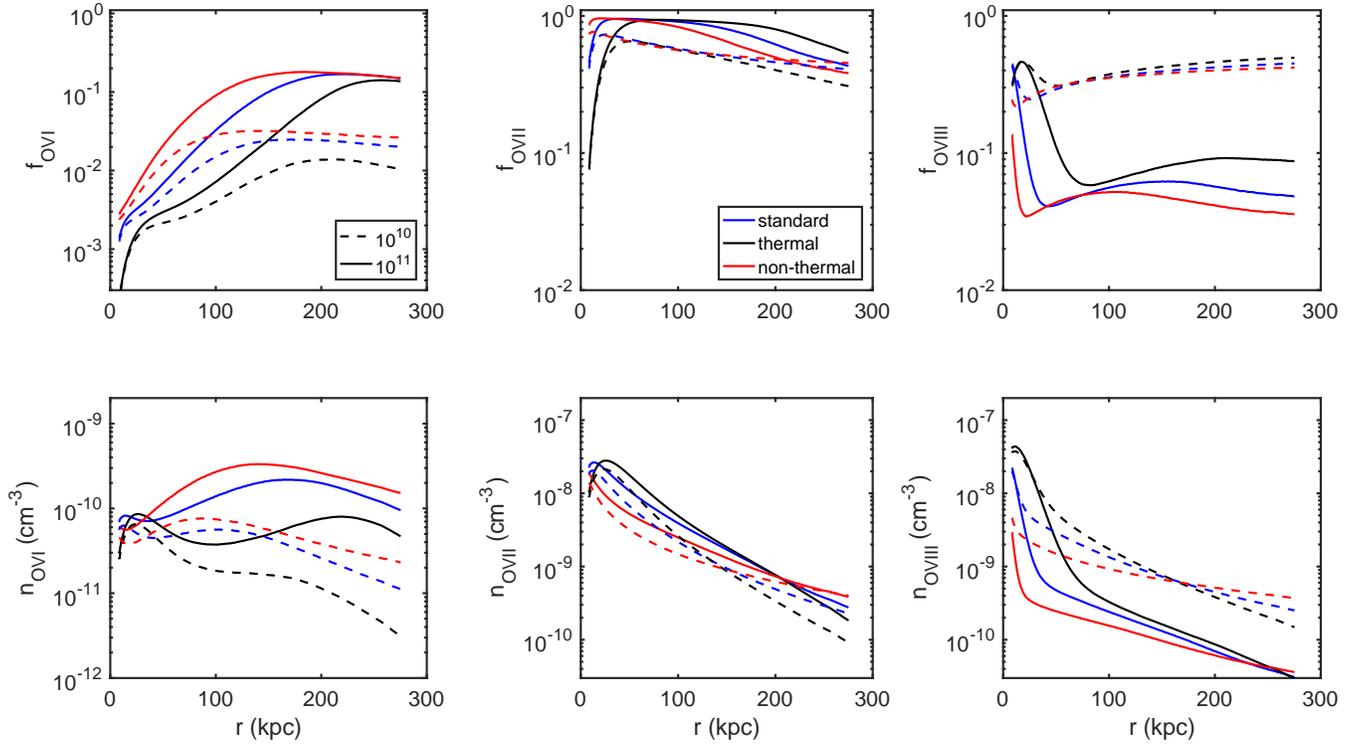}
 \caption{Oxygen ion fractions (top) and ion volume densities (bottom), for \ovi~(left), \ovii~(middle) and \oviii~(right).}
  \label{fig:app2}
\end{figure*}

Figure~\ref{fig:app2} shows the ion fractions and volume densities of the \ovi-\oviii~ions as functions of radius. The top panels show the oxygen ion fractions. The high \mgas~models (solid curves) show the effect of the different temperature profiles on the ion fractions. The low mass models (dashed) have low gas densities and are useful to examine the effect of photoionization (PI).

In the high mass models, the \ovi~fraction (left panel) at the outer boundary depends on the gas temperature, and it is identical in all three models. As we move inwards, the gas temperature increases and the \ovi~fraction decreases. It does so faster in the steep model, due to the steeper temperature profile (see Figure~\ref{fig:app1}). In the low mass models, gas densities are $<5 \times 10^{-5}$~\cmv~at $r>30$~kpc, low enough for PI to reduce the \ovi~fraction. The effect is stronger for the thermal model, where the gas density at large radii is lower. At small radii ($r\sim 10$~kpc), the gas densities are high enough for the ion fraction to determined by the gas temperature only (CIE) and they are identical for the low and high CGM models.

The \ovii~fraction (middle panel) is of order unity for most radii in all three density profiles, both for the high and low CGM mass models. Differences between the high mass models can be seen at very small radii, where the gas temperature in the steep models is high and \ovii~is collisionally ionized (CI) to form \oviii. At large radii, the \ovii~is higher for the thermal profile since some of the \ovi~is photoionized into \ovii. In the low mass models, some of the \ovii~is photoionized to \oviii, and the \ovii~fractions are overall lower at $r \gtrsim 30$~kpc. In the innermost region, $r<20$~kpc, the gas density is high enough for the gas to be in CIE, and the \ovii~fractions are similar for the low and high mass models.

The \oviii~(right panel) in the high mass models is formed by CI in the central parts, and the ion fraction is highest in the thermal model, where the temperature is optimal for \oviii, with $T_{\rm peak} \sim 2.5 \times 10^6$~K in CIE. Outside the CI core, the ion is formed by PI of the \ovii, and its fraction is low ($f_{\rm \oviii}<0.1$). In the low mass models, the lower gas densities lead to a significant increase in the \oviii~fraction, to $f_{\rm \oviii} \sim 0.3-0.4$. This is comparable to the peak CI fraction, and the ion fraction profiles are similar all the way to the central part of the halo.

The bottom panels show the ion volume densities, $n_{ion}(r) = n_HZ'f_{ion}$, and are useful for understanding the column density plots presented in Figure~\ref{fig:res_oxy_p2}. The \ovi~volume density (left) is lower in the thermal models at both low and high CGM masses, leading to lower column densities compared to the non-thermal models. As shown earlier, for high gas mass models this is a result of the gas temperature profile (CI), and for low mass models - the gas density (PI). The \ovi~ion fraction profiles increase with radius and offset the decline in density and metallicity, leading to ion volume densities that are almost flat or relatively slowly decreasing with radius, and large length scales (see bottom panels of Figure~\ref{fig:res_oxy_p2}).

The \ovii~and \oviii~profiles (middle and right panels), on the other hand, decrease rapidly for all models, and most of the column density forms in the inner part of the halo. The ion densities in the thermal models are higher out to $\sim 150$~kpc, resulting in higher column densities. Finally, the profiles of the high mass models intersect at larger radii than the low mass profiles, leading to larger (integrated) differences between the profile shapes.


\renewcommand{\theequation}{\arabic{equation}}

\end{document}